arXiv:1802.05918v1 [cs.PL] 16 Feb 2018

# Compositional Verification of Compiler Optimisations on Relaxed Memory

Mike Dodds[1], Mark Batty[2], and Alexey Gotsman[3]

[1] Galois Inc. [2] University of Kent [3] IMDEA Software Institute

**Abstract.** A valid compiler optimisation transforms a block in a program without introducing new observable behaviours to the program as a whole. Deciding which optimisations are valid can be difficult, and depends closely on the semantic model of the programming language. Axiomatic relaxed models, such as C++11, present particular challenges for determining validity, because such models allow subtle effects of a block transformation to be observed by the rest of the program. In this paper we present a denotational theory that captures optimisation validity on an axiomatic model corresponding to a fragment of C++11. Our theory allows verifying an optimisation compositionally, by considering only the block it transforms instead of the whole program. Using this property, we realise the theory in the first push-button tool that can verify real-world optimisations under an axiomatic memory model.

## 1 Introduction

*Context and objectives.* Any program defines a collection of observable behaviours: a sorting algorithm maps unsorted to sorted sequences, and a paint program responds to mouse clicks by updating a rendering. It is often desirable to transform a program without introducing new observable behaviours – for example, in a compiler optimisation or programmer refactoring. Such transformations are called *observational refinements*, and they ensure that properties of the original program will carry over to the transformed version. It is also desirable for transformations to be *compositional*, meaning that they can be applied to a block of code irrespective of the surrounding program context. Compositional transformations are particularly useful for automated systems such as compilers, where they are known as *peephole optimisations*.

The semantics of the language is highly significant in determining which transformations are valid, because it determines the ways that a block of code being transformed can interact with its context and thereby affect the observable behaviour of the whole program. Our work applies to a relaxed memory concurrent setting. Thus, the context of a code-block includes both code sequentially before and after the block, and code that runs in parallel. Relaxed memory means that different threads can observe different, apparently contradictory orders of events – such behaviour is permitted by programming languages to reflect CPU-level relaxations and to allow compiler optimisations.

We focus on *axiomatic* memory models of the type used in C/C++ and Java. In axiomatic models, program executions are represented by structures of memory actions and relations on them, and program semantics is defined by a set of axioms constraining these structures. Reasoning about the correctness of program transformations on such memory models is very challenging, and indeed, compiler optimisations have been repeatedly shown unsound with respect to models they were intended to support [25, 23]. The fundamental difficulty is that axiomatic models are defined in a global, non-compositional way, making it very challenging to reason compositionally about the single code-block being transformed.

*Approach.* Suppose we have a code-block $B$, embedded into an unknown program context. We define a *denotation* for the code-block which summarises its behaviour in a restricted representative context. The denotation consists of a set of *histories* which track interactions across the boundary between the code-block and its context, but abstract from internal structure of the code-block. We can then validate a transformation from code-block $B$ to $B'$ by comparing their denotations. This approach is compositional: it requires reasoning only about the code-blocks and representative contexts; the validity of the transformation in an arbitrary context will follow. It is also *fully abstract*, meaning that it can verify any valid transformation: considering only representative contexts and histories does not lose generality.

We also define a variant of our denotation that is *finite* at the cost of losing full abstraction. We achieve this by further restricting the form of contexts one needs to consider in exchange for tracking more information in histories. For example, it is unnecessary to consider executions where two context operations read from the same write.

Using this finite denotation, we implement a prototype verification tool, Stellite. Our tool converts an input transformation into a model in the Alloy language [11], and then checks that the transformation is valid using the Alloy* solver [17]. Our tool can prove or disprove a range of introduction, elimination, and exchange compiler optimisations. Many of these were verified by hand in previous work; our tool verifies them automatically.

*Contributions.* Our contribution is twofold. First, we define the first fully abstract denotational semantics for an axiomatic relaxed model. Previous proposals in this space targeted either non-relaxed sequential consistency [6] or much more restrictive operational relaxed models [7, 12, 20]. Second, we show it is feasible to automatically verify relaxed-memory program transformations. Previous techniques required laborious proofs by hand or in a proof assistant [24, 26, 27, 25, 23]. Our target model is derived from the C/C++ 2011 standard [21]. However, our aim is not to handle C/C++ per se (especially as the model is in flux in several respects; see §3.7). Rather we target the simplest axiomatic model rich enough to demonstrate our approach.

## 2 Observation and Transformation

*Observational refinement.* The notion of *observation* is crucial when determining how different programs are related. For example, observations might be I/O behaviour or writes to special variables. Given program executions $X_1$ and $X_2$, we write $X_1 \preccurlyeq_{\mathsf{ex}} X_2$ if the observations in $X_1$ are replicated in $X_2$ (defined formally in the following). Lifting this notion, a program $P_1$ *observationally refines* another $P_2$ if every observable behaviour of one could also occur with the other – we write this $P_1 \preccurlyeq_{\mathsf{pr}} P_2$. More formally, let $[\![-]\!]$ be the map from programs to sets of executions. Then we define $\preccurlyeq_{\mathsf{pr}}$ as:

$$P_1 \preccurlyeq_{\mathsf{pr}} P_2 \quad \stackrel{\Delta}{\iff} \quad \forall X_1 \in [\![P_1]\!].\, \exists X_2 \in [\![P_2]\!].\, X_1 \preccurlyeq_{\mathsf{ex}} X_2 \tag{1}$$

*Compositional transformation.* Many common program transformations are *compositional*: they modify a sequential fragment of the program without examining the rest of the program. We call the former the *code-block* and the latter its *context*. Contexts can include sequential code before and after the block, and concurrent code that runs in parallel with it. Code-blocks are sequential, i.e. they do not feature internal concurrency. A context $C$ and code-block $B$ can be composed to give a whole program $C(B)$.

A transformation $B_2 \rightsquigarrow B_1$ replaces some instance of the code-block $B_2$ with $B_1$. To validate such a transformation, we must establish whether *every* whole program containing $B_1$ observationally refines the same program with $B_2$ substituted. If this holds, we say that $B_1$ observationally refines $B_2$, written $B_1 \preccurlyeq_{\mathsf{bl}} B_2$, defined by lifting $\preccurlyeq_{\mathsf{pr}}$ as follows:

$$B_1 \preccurlyeq_{\mathsf{bl}} B_2 \quad \stackrel{\Delta}{\iff} \quad \forall C.\ C(B_1) \preccurlyeq_{\mathsf{pr}} C(B_2) \tag{2}$$

If $B_1 \preccurlyeq_{\mathsf{bl}} B_2$ holds, then the compiler can replace block $B_2$ with block $B_1$ irrespective of the whole program, i.e. $B_2 \rightsquigarrow B_1$ is a valid transformation. Thus, deciding $B_1 \preccurlyeq_{\mathsf{bl}} B_2$ is the core problem in validating compositional transformations.

The language semantics is highly significant in determining observational refinement. For example, the code blocks $B_1$: `store(x,5)` and $B_2$: `store(x,2); store(x,5)` are observationally equivalent in a sequential setting. However, in a concurrent setting the intermediate state, $\mathtt{x} = 2$, can be observed in $B_2$ but not $B_1$, meaning the code-blocks are no longer observationally equivalent. In a relaxed-memory setting there is no global state seen by all threads, which further complicates the notion of observation.

*Compositional verification.* To establish $B_1 \preccurlyeq_{\mathsf{bl}} B_2$, it is difficult to examine all possible syntactic contexts. Our approach is to construct a *denotation* for each code-block – a simplified, ideally finite, summary of possible interactions between the block and its context. We then define a *refinement relation* on denotations and use it to establish observational refinement. We write $B_1 \sqsubseteq B_2$ when the denotation of $B_1$ refines $B_2$.

Refinement on denotations should be *adequate*, i.e., it should validly approximate observational refinement: $B_1 \sqsubseteq B_2 \implies B_1 \preccurlyeq_{\mathsf{bl}} B_2$. Hence, if $B_1 \sqsubseteq B_2$, then $B_2 \rightsquigarrow B_1$ is a valid transformation. It is also desirable for the denotation to be *fully abstract*: $B_1 \preccurlyeq_{\mathsf{bl}} B_2 \implies B_1 \sqsubseteq B_2$. This means any valid transformation can be verified by comparing denotations. Below we define several versions of $\sqsubseteq$ with different properties.

## 3 Target Language and Core Memory Model

Our language's memory model is derived from the C/C++ 2011 standard (henceforth '*C11*'), as formalised by [21, 5]. However, we simplify our model in several ways; see

```
      store(x,0); store(y,0);
store(x,1);      || store(y,1);
v1 := load(y);   || v2 := load(x);
```

```
      store(f,0); store(x,0);
store(x,1);  || b := load(f);
store(f,1);  || if (b == 1)
             ||   r := load(x);
```

**Fig. 1.** *Left:* store-buffering (SB) example. *Right:* message-passing (MP) example.

the end of section for details. In C11 terms, our model covers release-acquire and non-atomic operations, and sequentially consistent fences. To simplify the presentation, at first we omit non-atomics, and extend our approach to cover them in §7. Thus, all operations in this section correspond to C11's release-acquire.

### 3.1 Relaxed Memory Primer

In a sequentially consistent concurrent system, there is a total temporal order on loads and stores, and loads take the value of the most recent store; in particular, they cannot read overwritten values, or values written in the future. A *relaxed* (or *weak*) memory model weakens this total order, allowing behaviours forbidden under sequential consistency. Two standard examples of relaxed behaviour are *store buffering (SB)* and *message passing (MP)*, shown in Figure 1.

In most relaxed models $\texttt{v1} = \texttt{v2} = 0$ is a possible post-state for SB. This cannot occur on a sequentially consistent system: if $\texttt{v1} = 0$, then `store(y,1)` must be ordered after the load of `y`, which would order `store(x,1)` before the load of `x`, forcing it to assign $\texttt{v2} = 1$. In some relaxed models, $\texttt{b} = 1 \wedge \texttt{r} = 0$ is a possible post-state for MP. This is undesirable if, for example, `x` is a complex data-structure and `f` is a flag indicating it has been safely created.

### 3.2 Language Syntax

Programs in the language we consider manipulate *thread-local variables* $l, l_1, l_2 \ldots \in \mathsf{LVar}$ and *global variables* $x, y, \ldots \in \mathsf{GVar}$, coming from disjoint sets $\mathsf{LVar}$ and $\mathsf{GVar}$. Each variable stores a value from a finite set $\mathsf{Val}$ and is initialised to $0 \in \mathsf{Val}$. Constants are encoded by special read-only thread-local variables. We assume that each thread uses the same set of thread-local variable names $\mathsf{LVar}$. The syntax of the programming language is as follows:

$$\begin{aligned} C \ ::=\ & l := E \mid \texttt{store}(x, l) \mid l := \texttt{load}(x) \mid l := \texttt{LL}(x) \mid l' := \texttt{SC}(x, l) \mid \texttt{fence} \mid \\ & C_1 \parallel C_2 \mid C_1; C_2 \mid \texttt{if}\,(l)\,\{C_1\}\,\texttt{else}\,\{C_2\} \mid \{-\} \\ E \ ::=\ & l \mid l_1 = l_2 \mid l_1 \neq l_2 \mid \ldots \end{aligned}$$

Many of the constructs are standard. $\texttt{LL}(x)$ and $\texttt{SC}(x, l)$ are *load-link* and *store-conditional*, which are basic concurrency operations available on many platforms (e.g., Power and ARM). A load-link $\texttt{LL}(x)$ behaves as a standard load of global variable `x`. However, if it is followed by a store-conditional $\texttt{SC}(x, l)$, the store fails and returns false if there are intervening writes to the same location. Otherwise the store-conditional writes $l$ and returns true. The `fence` command is a *sequentially consistent fence*: interleaving such fences between all statements in a program guarantees sequentially consistent behaviour. We do not include *compare-and-swap* (CAS) command in our language because LL-SC is more general [2]. Hardware-level LL-SC is used to implement C11 CAS on Power and ARM. Our language does not include loops because our model in this paper does not include infinite computations (see §3.7 for discussion). As a result, loops can be represented by their finite unrollings. Our `load` commands write into a local variable. In examples, we sometimes use 'bare' loads without a variable write.

The construct $\{-\}$ represents a block-shaped hole in the program. To simplify our presentation, we assume that at most one hole appears in the program. Transformations that apply to multiple blocks at once can be simulated by using the fact our approach is compositional: transformations can be applied in sequence using different divisions of the program into code-block and context.

The set $\mathsf{Prog}$ of *whole programs* consists of programs without holes, while the set $\mathsf{Contx}$ of *contexts* consists of programs with a hole. The set $\mathsf{Block}$ of *code-blocks* are whole programs without parallel composition. We often write $P \in \mathsf{Prog}$ for a whole program, $B \in \mathsf{Block}$ for a code-block, and $C \in \mathsf{Contx}$ for a context. Given a context $C$ and a code-block $B$, the composition $C(B)$ is $C$ with its hole syntactically replaced by $B$. For example:

$C$: `load(x); {-}; store(y,l1),` $B$: `store(x,2)`
$\longrightarrow$ $C(B)$: `load(x); store(x,2); store(y,l1)`

We restrict $\mathsf{Prog}$, $\mathsf{Contx}$ and $\mathsf{Block}$ to ensure LL-SC pairs are matched correctly. Each SC must be preceded in program order by a LL to the same location. Other types of operations may occur between the LL and SC, but intervening SC operations are forbidden. For example, the program `LL(x); SC(x,v1); SC(x,v2);` is forbidden. We also forbid LL-SC pairs from spanning parallel compositions, and from spanning the block/context boundary.

### 3.3 Memory Model Structure

The semantics of a whole program $P$ is given by a set $[\![P]\!]$ of *executions*, which consist of *actions*, representing memory events on global variables, and several relations on these. Actions are tuples in the set $\mathsf{Action} \triangleq \mathsf{ActID} \times \mathsf{Kind} \times \mathsf{Option}(\mathsf{GVar}) \times \mathsf{Val}^*$. In an action $(a, k, z, b) \in \mathsf{Action}$: $a \in \mathsf{ActID}$ is the unique action identifier; $k \in \mathsf{Kind}$ is the kind of action – we use $\mathsf{load}$, $\mathsf{store}$, $\mathsf{LL}$, $\mathsf{SC}$, and the failed variant $\mathsf{SC}_f$ in the semantics, and will introduce further kinds as needed; $z \in \mathsf{Option}(\mathsf{GVar})$ is an option type consisting of either a single global variable $\mathsf{Just}(x)$ or $\mathsf{None}$; and $b \in \mathsf{Val}^*$ is the vector of values (actions with multiple values are used in §4).

Given an action $v$, we use $\mathsf{gvar}(v)$ and $\mathsf{val}(v)$ as selectors for the different fields. We often write actions so as to elide action identifiers and the option type. For example, $\mathsf{load}(x, 3)$ stands for $\exists i.\, (i, \mathsf{load}, \mathsf{Just}(x), [3])$. We also sometimes elide values. We call $\mathsf{load}$ and $\mathsf{LL}$ actions *reads*, and $\mathsf{store}$ and successful $\mathsf{SC}$ actions *writes*. Given a set of actions $\mathcal{A}$, we write, e.g., $\mathsf{reads}(\mathcal{A})$ to identify read actions in $\mathcal{A}$. Below, we range over all actions by $u, v$; read actions by $r$; write actions by $w$; and $\mathsf{LL}$, $\mathsf{SC}$ actions by *ll* and *sc* respectively.

$$
\begin{aligned}
\langle l := \texttt{load}(x), \sigma\rangle &\triangleq \{(\{\mathsf{load}(x,a)\}, \emptyset, \sigma[l \mapsto a]) \mid a \in \mathsf{Val}\}\\
\langle \texttt{store}(x,l), \sigma\rangle &\triangleq \{(\{\mathsf{store}(x,a)\}, \emptyset, \sigma) \mid \sigma(l) = a\}\\
\langle C_1; C_2, \sigma\rangle &\triangleq \{(\mathcal{A}_1 \uplus \mathcal{A}_2, \mathsf{sb}_1 \cup \mathsf{sb}_2 \cup (\mathcal{A}_1 \times \mathcal{A}_2), \sigma_2) \mid\\
&\qquad (\mathcal{A}_1, \mathsf{sb}_1, \sigma_1) \in \langle C_1, \sigma\rangle \wedge (\mathcal{A}_2, \mathsf{sb}_2, \sigma_2) \in \langle C_2, \sigma_1\rangle\}\\
\langle \texttt{fence}, \sigma\rangle &\triangleq \{(\{\mathit{ll}, \mathit{sc}\}, \{(\mathit{ll}, \mathit{sc})\}, \sigma) \mid \mathit{ll} = \mathsf{LL}(\mathit{fen}, 0) \wedge \mathit{sc} = \mathsf{SC}(\mathit{fen}, 0)\}
\end{aligned}
$$

**Fig. 2.** Selected clauses of the thread-local semantics. The full semantics is given in §A. We write $\mathcal{A}_1 \uplus \mathcal{A}_2$ for a union that is defined only when actions in $\mathcal{A}_1$ and $\mathcal{A}_2$ use disjoint sets of identifiers. We omit identifiers from actions to avoid clutter.

The semantics of a program $P \in \mathsf{Prog}$ is defined in two stages. First, a *thread-local semantics* of $P$ produces a set $\langle P\rangle$ of *pre-executions* $(\mathcal{A}, \mathsf{sb}) \in \mathsf{PreExec}$. A pre-execution contains a finite set of memory actions $\mathcal{A} \subseteq \mathsf{Action}$ that could be produced by the program. It has a transitive and irreflexive *sequence-before* relation $\mathsf{sb} \subseteq \mathcal{A} \times \mathcal{A}$, which defines the sequential order imposed by the program syntax.

For example two sequential statements in the same thread produce actions ordered in $\mathsf{sb}$. The thread-local semantics takes into account control flow in $P$'s threads and operations on local variables. However, it does not constrain the behaviour of global variables: the values threads read from them are chosen arbitrarily. This is addressed by extending pre-executions with extra relations, and filtering the resulting *executions* using *validity axioms*.

### 3.4 Thread-Local Semantics

The thread-local semantics is defined formally in Figure 2. The semantics of a program $P \in \mathsf{Prog}$ is defined using function $\langle -, -\rangle\colon \mathsf{Prog} \times \mathsf{VMap} \to \mathcal{P}(\mathsf{PreExec} \times \mathsf{VMap})$. The values of local variables are tracked by a map $\sigma \in \mathsf{VMap} \triangleq \mathsf{LVar} \to \mathsf{Val}$. Given a program and an input local variable map, the function produces a set of pre-executions paired with an output variable map, representing the values of local variables at the end of the execution. Let $\sigma_0$ map every local variable to $0$. Then $\langle P\rangle$, the thread-local semantics of a program $P$, is defined as

$$\langle P\rangle \quad \triangleq \quad \{(\mathcal{A}, \mathsf{sb}) \mid \exists \sigma'.\, (\mathcal{A}, \mathsf{sb}, \sigma') \in \langle P, \sigma_0\rangle\}$$

The significant property of the thread-local semantics is that it does not restrict the behaviour of global variables. For this reason, note that the clause for `load` in Figure 2 leaves the value $a$ unrestricted. We follow [15] in encoding the `fence` command by a successful LL-SC pair to a distinguished variable *fen* $\in \mathsf{GVar}$ that is not otherwise read or written.

### 3.5 Execution Structure and Validity Axioms

The semantics of a program $P$ is a set $[\![P]\!]$ of *executions* $X = (\mathcal{A}, \mathsf{sb}, \mathsf{at}, \mathsf{rf}, \mathsf{mo}, \mathsf{hb}) \in \mathsf{Exec}$, where $(\mathcal{A}, \mathsf{sb})$ is a pre-execution and $\mathsf{at}, \mathsf{rf}, \mathsf{mo}, \mathsf{hb} \subseteq \mathcal{A} \times \mathcal{A}$. Given an execution $X$ we sometimes write $\mathcal{A}(X), \mathsf{sb}(X), \ldots$ as selectors for the appropriate set or relation. The relations have the following purposes.

- *Reads-from* (rf) is an injective map from reads to writes at the same location of the same value. A read and a write actions are related $w \xrightarrow{\mathsf{rf}} r$ if $r$ takes its value from $w$.
- *Modification order* (mo) is an irreflexive, total order on write actions to each distinct variable. This is a per-variable order in which *all* threads observe writes to the variable; two threads cannot observe these writes in different orders.
- *Happens-before* (hb) is analogous to global temporal order – but unlike the sequentially consistent notion of time, it is partial. Happens-before is defined as $(\mathsf{sb} \cup \mathsf{rf})^+$: therefore statements ordered in the program syntax are ordered in time, as are reads with the writes they observe.
- *Atomicity* ($\mathsf{at} \subseteq \mathsf{sb}$) is an extension to standard C11 which we use to support LL-SC (see below). It is an injective function from a successful load-link action to a successful store-conditional, giving a LL-SC pair.

The semantics $[\![P]\!]$ of a program $P$ is the set of executions $X \in \mathsf{Exec}$ compatible with the thread-local semantics and the *validity axioms*, denoted $\mathsf{valid}(X)$:

$$[\![P]\!] \triangleq \{X \mid (\mathcal{A}(X), \mathsf{sb}(X)) \in \langle P \rangle \land \mathsf{valid}(X)\} \tag{3}$$

The validity axioms on an execution $(\mathcal{A}, \mathsf{sb}, \mathsf{at}, \mathsf{rf}, \mathsf{mo}, \mathsf{hb})$ are:

- HBDEF: $\mathsf{hb} = (\mathsf{sb} \cup \mathsf{rf})^+$ *and* hb is acyclic.
  This axiom defines hb and enforces the intuitive property that there are no cycles in the temporal order. It also prevents an action reading from its hb-future: as rf is included in hb, this would result in a cycle.
- HBVSMO: $\neg\exists w_1, w_2.\ w_1 \underset{\mathsf{mo}}{\overset{\mathsf{hb}}{\rightleftarrows}} w_2$
  This axiom requires that the order in which writes to a location become visible to threads cannot contradict the temporal order. But take note that writes may be ordered in mo but not hb.
- COHERENCE: $\neg\exists w_1, w_2, r.\ w_1 \xrightarrow{\mathsf{mo}} w_2 \xrightarrow{\mathsf{hb}} r \land w_1 \xrightarrow{\mathsf{rf}} r$
  This axiom generalises the sequentially consistent prohibition on reading overwritten values. If two writes are ordered in mo, then intuitively the second overwrites the first. A read that follows some write in hb or mo cannot read from writes earlier in mo – these earlier writes have been overwritten. However, unlike in sequential consistency, hb is partial, so there may be multiple writes that an action can legally read.
- RFVAL: $\forall r. (\neg\exists w'. w' \xrightarrow{\mathsf{rf}} r) \implies (\mathsf{val}(r) = 0 \land (\neg\exists w. w \xrightarrow{\mathsf{hb}} r \land \mathsf{gvar}(w) = \mathsf{gvar}(r)))$
  Most reads must take their value from a write, represented by an rf edge. However, the RFVAL axiom allows the rf edge to be omitted if the read takes the initial value 0 and there is no hb-earlier write to the same location. Intuitively, an hb-earlier write would supersede the initial value in a similar way to COHERENCE.

– ATOM: $\neg\exists w_1, w_2, ll, sc.$

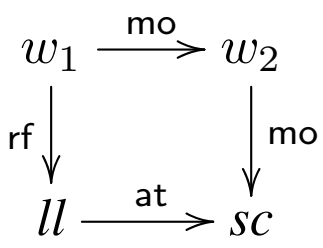

This axiom is adapted from [15]. For an LL-SC pair *ll* and *sc*, it ensures that there is no mo-intervening write $w_2$ that would invalidate the store.

Our model forbids the problematic relaxed behaviour of the message-passing (MP) program in Figure 1 that yields $\mathtt{b} = 1 \wedge \mathtt{r} = 0$. Figure 3 shows an (invalid) execution that would exhibit this behaviour. To avoid clutter, here and in the following we omit hb edges obtained by transitivity and local variable values. This execution is allowed by the thread-local semantics of the MP program, but it is ruled out by the COHERENCE validity axiom. As hb is transitively closed, there is a derived hb edge $\mathsf{store}(\mathtt{x}, 1) \xrightarrow{\mathsf{hb}} \mathsf{load}(\mathtt{x}, 0)$, which forms a COHERENCE violation. Thus, this is not an execution of the MP program. Indeed, any execution ending in $\mathsf{load}(\mathtt{x}, 0)$ is forbidden for the same reason, meaning that the MP relaxed behaviour cannot occur.

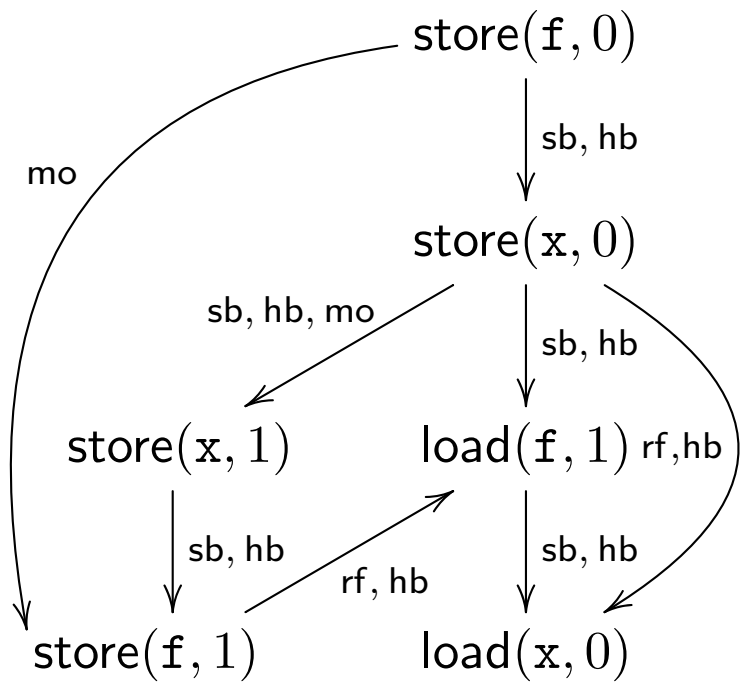

**Fig. 3.** An invalid execution of MP.

### 3.6 Relaxed Observations

Finally, we define a notion of observational refinement suitable for our relaxed model. We assume a subset of *observable* global variables, $\mathsf{OVar} \subseteq \mathsf{GVar}$, which can only be accessed by the context and not by the code-block. We consider the actions and the hb relation on these variables to be the observations. We write $X|_{\mathsf{OVar}}$ for the projection of $X$'s action set and relations to OVar, and use this to define $\preccurlyeq_{\mathsf{ex}}$ for our model:

$$X \preccurlyeq_{\mathsf{ex}} Y \quad \overset{\Delta}{\iff} \quad \mathcal{A}(X|_{\mathsf{OVar}}) = \mathcal{A}(Y|_{\mathsf{OVar}}) \wedge \mathsf{hb}(Y|_{\mathsf{OVar}}) \subseteq \mathsf{hb}(X|_{\mathsf{OVar}})$$

This is lifted to programs and blocks as in §2, def. (1) and (2). Note that in the more abstract execution, actions on observable variables must be the same, but hb can be weaker. This is because we interpret hb as a constraint on time order: two actions that are unordered in hb could have occurred in either order, or in parallel. Thus, weakening hb allows more observable behaviours (see §2).

### 3.7 Differences from C11

Our language's memory model is derived from the C11 formalisation in [5], with a number of simplifications. We chose C11 because it demonstrates most of the important features of axiomatic language models. However, we do not target the precise C11 model: rather we target an abstracted model that is rich enough to demonstrate our approach. Relaxed language semantics is still a very active topic of research, and several

C11 features are known to be significantly flawed, with multiple competing fixes proposed. Some of our differences from [5] are intended to avoid such problematic features so that we can cleanly demonstrate our approach.

In C11 terms, our model covers release-acquire and non-atomic operations (the latter addressed in §7), and sequentially consistent fences. We deviate from C11 in the following ways:

- We omit *sequentially consistent* accesses because their semantics is known to be flawed in C11 [16]. We do handle sequentially consistent fences, but these are stronger than those of C11: we use the semantics proposed in [15]. It has been proved sound under existing compilation strategies to common multiprocessors.
- We omit *relaxed* (RLX) accesses to avoid well-known problems with thin-air values [4]. There are multiple recent competing proposals for fixing these problems, e.g. [14, 13, 19].
- Our model does not include infinite computations, because their semantics in C11-style axiomatic models remains undecided in the literature [4]. However, our proofs do not depend on the assumption that execution contexts are finite.
- Our language is based on shared variables, not dynamically allocated addressable memory, so for example we cannot write `y:=*x; z:=*y`. This simplifies our theory by allowing us to fix the variables accessed by a code-block up-front. We believe our results can be extended to support addressable memory, because C11-style models grant no special status to pointers; we elaborate on this in §4.
- We add LL-SC atomic instructions to our language in addition to C11's standard CAS. To do this, we adapt the approach of [15]. This increases the observational power of a context and is necessary for full abstraction in the presence of non-atomics; see §8. LL-SC is available as a hardware instruction on many platforms supporting C11, such as Power and ARM. However, we do not propose adding LL-SC to C11: rather, it supports an interesting result in relaxed memory model theory. Our adequacy results do not depend on LL-SC.

# 4 Denotations of Code-Blocks

We construct the denotation for a code-block in two steps: (1) generate the *block-local* executions under a set of special cut-down contexts; (2) from each execution, extract a summary of interactions between the code-block and the context called a *history*.

## 4.1 Block-Local Executions

The block-local executions of a block $B \in \mathsf{Block}$ omit context structure such as syntax and actions on variables not accessed in the block. Instead the context is represented by special actions $\mathsf{call}$ and $\mathsf{ret}$, a set $\mathcal{A}_B$, and relations $R_B$ and $S_B$, each covering an aspect of the interaction of the block and an arbitrary unrestricted context. Together, each choice of $\mathsf{call}$, $\mathsf{ret}$, $\mathcal{A}_B$, $R_B$, and $S_B$ abstractly represents a set of possible syntactic contexts. By quantifying over the possible values of these parameters, we cover the behaviour of *all* syntactic contexts. The parameters are defined as follows:

- *Local variables.* A context can include code that precedes and follows the block on the same thread, with interaction through local variables, but – due to syntactic restriction – not through LL/SC atomic regions. We capture this with special action $\mathsf{call}(\sigma)$ at the start of the block, and $\mathsf{ret}(\sigma')$ at the end, where $\sigma, \sigma' \colon \mathsf{LVar} \rightarrow \mathsf{Val}$ record the values of local variables at these points. Assume that variables in LVar are ordered: $l_1, l_2, \ldots, l_n$. Then $\mathsf{call}(\sigma)$ is encoded by the action $(i, \mathsf{call}, \mathsf{None}, [\sigma(l_1), \ldots \sigma(l_n)])$, with fresh identifier $i$. We encode ret in the same way.
- *Global variable actions.* The context can also interact with the block through concurrent reads and writes to global variables. These interactions are represented by set $\mathcal{A}_B$ of *context actions* added to the ones generated by the thread-local semantics of the block. This set only contains actions on the variables $\mathsf{VS}_B$ that $B$ can access ($\mathsf{VS}_B$ can be constructed syntactically). Given an execution $X$ constructed using $\mathcal{A}_B$ (see below) we write $\mathsf{contx}(X)$ to recover the set $\mathcal{A}_B$.
- *Context happens-before.* The context can generate hb edges between its actions, which affect the behaviour of the block. We track these effects with a relation $R_B$ over actions in $\mathcal{A}_B$, call and ret:

$$R_B \subseteq (\mathcal{A}_B \times \mathcal{A}_B) \cup (\mathcal{A}_B \times \{\mathsf{call}\}) \cup (\{\mathsf{ret}\} \times \mathcal{A}_B) \quad (4)$$

  The context can generate hb edges between actions directly if they are on the same thread, or indirectly through inter-thread reads. Likewise call / ret may be related to context actions on the same or different threads.
- *Context atomicity.* The context can generate at edges between its actions that we capture in the relation $S_B \subseteq \mathcal{A}_B \times \mathcal{A}_B$. We require this relation to be an injective function from LL to SC actions. We consider only cases where LL/SC pairs do not cross block boundaries, so we need not consider boundary-crossing at edges.

Together, call, ret, $\mathcal{A}_B$, $R_B$, and $S_B$ represent a limited context, stripped of syntax, relations sb, mo, and rf, and actions on global variables other than $\mathsf{VS}_B$. When constructing block-local executions, we represent all possible interactions by quantifying over all possible choices of $\sigma$, $\sigma'$, $\mathcal{A}_B$, $R_B$ and $S_B$. The set $[\![B, \mathcal{A}_B, R_B, S_B]\!]$ contains all executions of $B$ under this special limited context. Formally, an execution $X = (\mathcal{A}, \mathsf{sb}, \mathsf{at}, \mathsf{rf}, \mathsf{mo}, \mathsf{hb})$ is in this set if:

1. $\mathcal{A}_B \subseteq \mathcal{A}$ and there exist variable maps $\sigma, \sigma'$ such that $\{\mathsf{call}(\sigma), \mathsf{ret}(\sigma')\} \subseteq \mathcal{A}$. That is, the call, return, and extra context actions are included in the execution.
2. There exists a set $\mathcal{A}_l$ and relation $\mathsf{sb}_l$ such that (i) $(\mathcal{A}_l, \mathsf{sb}_l, \sigma') \in \langle B, \sigma \rangle$; (ii) $\mathcal{A}_l = \mathcal{A} \setminus (\mathcal{A}_B \cup \{\mathsf{call}, \mathsf{ret}\})$; (iii) $\mathsf{sb}_l = \mathsf{sb} \setminus \{(\mathsf{call}, u), (u, \mathsf{ret}) \mid u \in \mathcal{A}_l\}$. That is, actions from the code-block satisfy the thread-local semantics, beginning with map $\sigma$ and deriving map $\sigma'$. All actions arising from the block are between call and ret in sb.
3. $X$ satisfies the validity axioms, but with modified axioms HBDEF′ and ATOM′. We define HBDEF′ as: $\mathsf{hb} = (\mathsf{sb} \cup \mathsf{rf} \cup R_B)^+$ and hb is acyclic. That is, context relation $R_B$ is added to hb. ATOM′ is defined analogously with $S_B$ added to at.

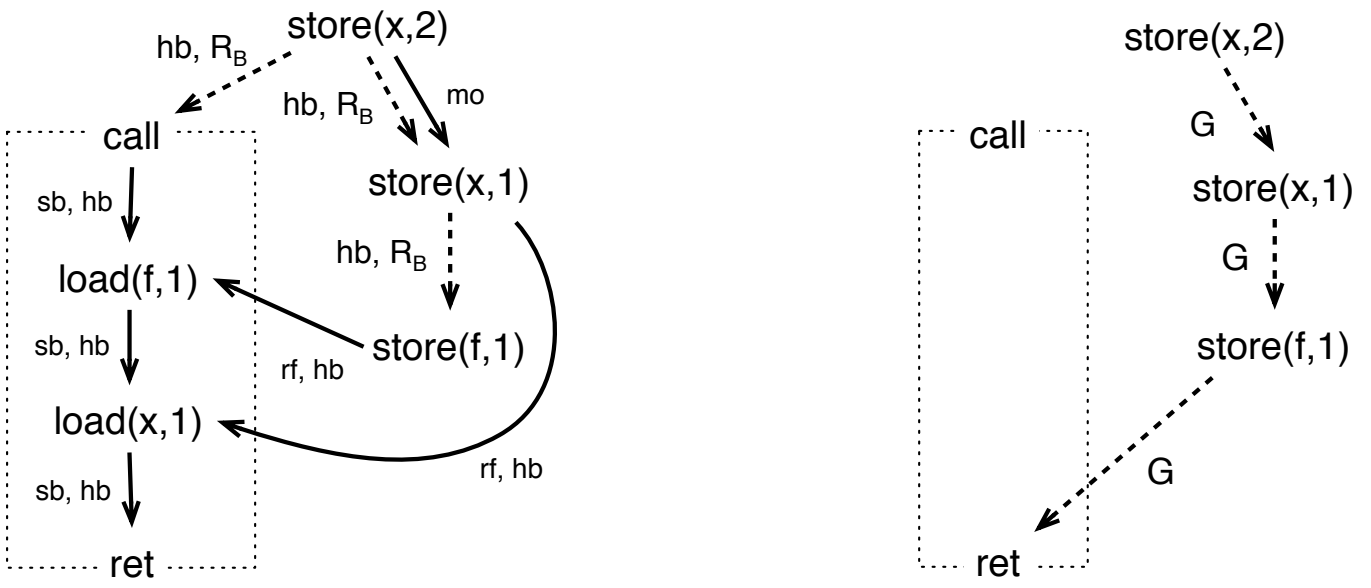

**Fig. 4.** *Left:* block-local execution. *Right:* corresponding history.

We say that $\mathcal{A}_B$, $R_B$ and $S_B$ are *consistent with* $B$ if they act over variables in the set $\mathsf{VS}_B$. In the rest of the paper we only consider consistent choices of $\mathcal{A}_B$, $R_B$, $S_B$. The *block-local executions* of $B$ are then all executions $X \in [\![B, \mathcal{A}_B, R_B, S_B]\!]$.[4]

*Example block-local execution.* The left of Figure 4 shows a block-local execution for the code-block

$$\texttt{l1 := load(f); l2 := load(x)} \tag{5}$$

Here the set $\mathsf{VS}_B$ of accessed global variables is $\{\texttt{f}, \texttt{x}\}$, As before, we omit local variables to avoid clutter. The context action set $\mathcal{A}_B$ consists of the three stores, and $R_B$ is denoted by dotted edges.

In this execution, both $\mathcal{A}_B$ and $R_B$ affect the behaviour of the code-block. The following path is generated by $R_B$ and the load of $\texttt{f} = 1$:

$$\mathsf{store}(\texttt{x}, 2) \xrightarrow{\mathsf{mo}} \mathsf{store}(\texttt{x}, 1) \xrightarrow{R_B} \mathsf{store}(\texttt{f}, 1) \xrightarrow{\mathsf{rf}} \mathsf{load}(\texttt{f}, 1) \xrightarrow{\mathsf{sb}} \mathsf{load}(\texttt{x}, 1)$$

Because $\mathsf{hb}$ includes $\mathsf{sb}$, $\mathsf{rf}$, and $R_B$, there is a transitive edge $\mathsf{store}(\texttt{x}, 1) \xrightarrow{\mathsf{hb}} \mathsf{load}(\texttt{x}, 1)$. The edge $\mathsf{store}(\texttt{x}, 2) \xrightarrow{\mathsf{mo}} \mathsf{store}(\texttt{x}, 1)$ is forced because the HBvsMO axiom prohibits $\mathsf{mo}$ from contradicting $\mathsf{hb}$. Consequently, the COHERENCE axiom forces the code-block to read $\texttt{x} = 1$.

### 4.2 Histories

From any block-local execution $X$, its *history* summarises the interactions between the code-block and the context. Informally, the history records $\mathsf{hb}$ over context actions, $\mathsf{call}$, and $\mathsf{ret}$. More formally the history, written $\mathsf{hist}(X)$, is a pair $(\mathcal{A}, G)$ consisting of an action set $\mathcal{A}$ and *guarantee relation* $G \subseteq \mathcal{A} \times \mathcal{A}$. Recall that we use $\mathsf{contx}(X)$ to denote the set of context actions in $X$. Using this, we define the history as follows:

- The action set $\mathcal{A}$ is the projection of $X$'s action set to call, ret, and $\mathsf{contx}(X)$.
- The guarantee relation $G$ is the projection of $\mathsf{hb}(X)$ to

$$(\mathsf{contx}(X) \times \mathsf{contx}(X)) \cup (\mathsf{contx}(X) \times \{\mathsf{ret}\}) \cup (\{\mathsf{call}\} \times \mathsf{contx}(X)) \qquad (6)$$

[4] This definition relies on the fact that our language supports a fixed set of global variables, not dynamically allocated addressable memory (see §3.7). We believe that in the future our results can be extended to support dynamic memory. For this, the block-local construction would need to quantify over actions on all possible memory locations, not just the static variable set $\mathsf{VS}_B$. The rest of our theory would remain the same, because C11-style models grant no special status to pointer values. Cutting down to a finite denotation, as in §5 below, would require some extra abstraction over memory – for example, a separation logic domain such as [9].

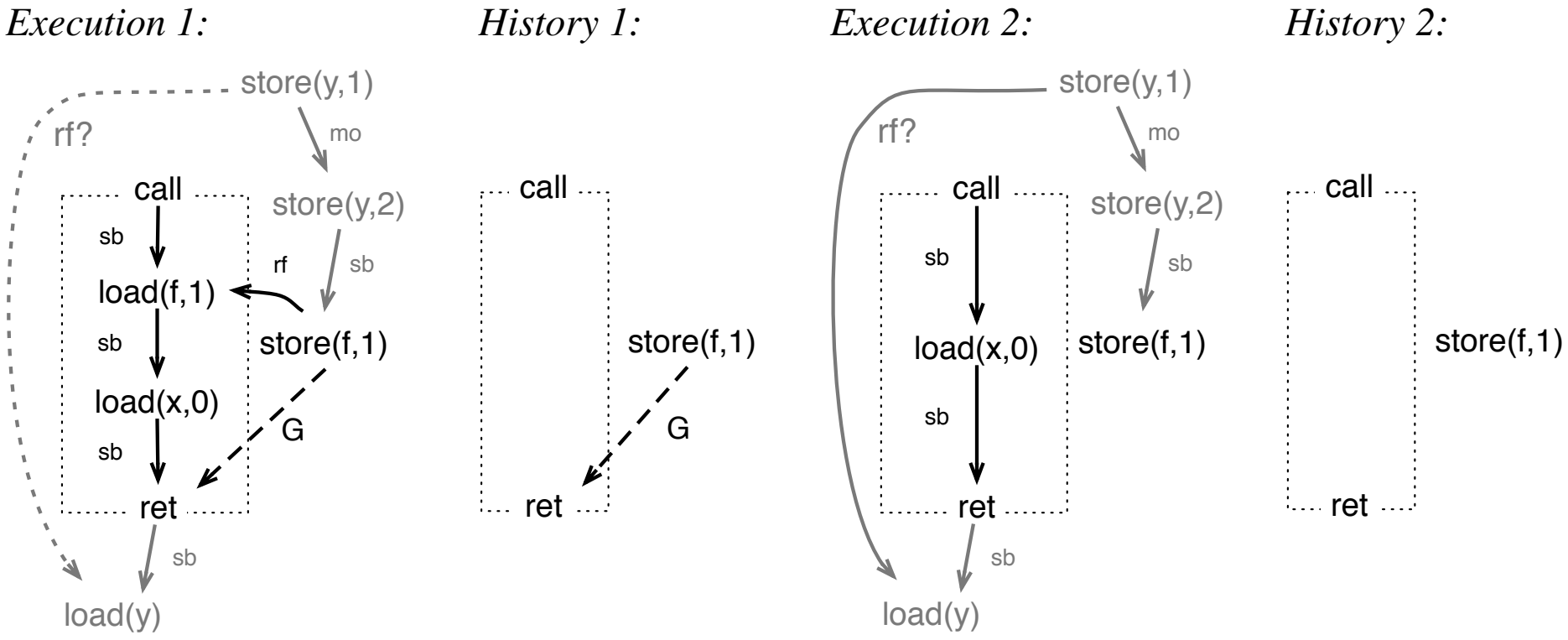

**Fig. 5.** Executions and histories illustrating the guarantee relation.

The guarantee summarises the code-block's effect on its context: it suffices to only track hb and ignore other relations. Note the guarantee definition is similar to the context relation $R_B$, definition (4). The difference is that call and ret are switched: this is because the guarantee represents hb edges generated by the code-block, while $R_B$ represents the edges generated by the context. The right of Figure 4 shows the history corresponding to the block-local execution on the left.

To see the interactions captured by the guarantee, compare the block given in def. (5) with the block `l2:=load(x)`. These blocks have differing effects on the following syntactic context:

```
store(y,1); store(y,2); store(f,1)   ||   {-}; l3:=load(y)
```

For the two-load block embedded into this context, $\texttt{l1} = 1 \wedge \texttt{l3} = 1$ is not a possible post-state. For the single-load block, this post-state is permitted.[5]

In Figure 4.2, we give executions for both blocks embedded into this context. We draw the context actions that are not included into the history in grey. In these executions, the code block determines whether the load of y can read value 1 (represented by the edge labelled 'rf?'). In the first execution, the context load of y cannot read 1 because there is the path $\mathsf{store}(\mathtt{y},1) \xrightarrow{\mathsf{mo}} \mathsf{store}(\mathtt{y},2) \xrightarrow{\mathsf{hb}} \mathsf{load}(\mathtt{y})$ which would contradict the COHERENCE axiom. In the second execution there is no such path and the load may read 1.

[5] We choose these post-states for exposition purposes – in fact these blocks are also distinguishable through local variable `l1` alone.

It is desirable for our denotation to hide the precise operations inside the block – this lets it relate syntactically distinct blocks. Nonetheless, the history must record $\mathsf{hb}$ effects such as those above that are visible to the context. In Execution 1, the COHERENCE violation is still visible if we only consider context operations, $\mathsf{call}$, $\mathsf{ret}$, and the guarantee $G$ – i.e. the history. In Execution 2, the fact that the read is permitted is likewise visible from examining the history. Thus the guarantee, combined with the local variable post-states, capture the effect of the block on the context without recording the actions inside the block.

### 4.3 Comparing Denotations

The denotation of a code-block $B$ is the set of histories of block-local executions of $B$ under each possible context, i.e. the set

$$\{\mathsf{hist}(X) \mid \exists \mathcal{A}_B, R_B, S_B.\, X \in [\![B, \mathcal{A}_B, R_B, S_B]\!]\}$$

To compare the denotations of two code-blocks, we first define a *refinement relation* on histories: $(\mathcal{A}_1, G_1) \sqsubseteq_{\mathsf{h}} (\mathcal{A}_2, G_2)$ holds iff $\mathcal{A}_1 = \mathcal{A}_2 \wedge G_2 \subseteq G_1$. The history $(\mathcal{A}_2, G_2)$ places fewer restrictions on the context than $(\mathcal{A}_1, G_1)$ – a weaker guarantee corresponds to more observable behaviours. For example in Figure 4.2, *History 1* $\sqsubseteq_{\mathsf{h}}$ *History 2* but not vice versa, which reflects the fact that History 1 rules out the read pattern discussed above.

We write $B_1 \sqsubseteq_{\mathsf{q}} B_2$ to state that the denotation of $B_1$ *refines* that of $B_2$. The subscript 'q' stands for the fact we *quantify* over both $\mathcal{A}$ and $R_B$. We define $\sqsubseteq_{\mathsf{q}}$ by lifting $\sqsubseteq_{\mathsf{h}}$:

$$B_1 \sqsubseteq_{\mathsf{q}} B_2 \overset{\Delta}{\iff} \forall \mathcal{A}, R, S.\, \forall X_1 \in [\![B_1, \mathcal{A}, R, S]\!].\; \exists X_2 \in [\![B_2, \mathcal{A}, R, S]\!].\, \mathsf{hist}(X_1) \sqsubseteq_{\mathsf{h}} \mathsf{hist}(X_2) \tag{7}$$

In other words, two code-blocks are related $B_1 \sqsubseteq_{\mathsf{q}} B_2$ if for every block-local execution of $B_1$, there is a corresponding execution of $B_2$ with a related history. Note that the corresponding history must be constructed under the same cut-down context $\mathcal{A}, R, S$.

THEOREM 1 (ADEQUACY OF $\sqsubseteq_{\mathsf{q}}$) $B_1 \sqsubseteq_{\mathsf{q}} B_2 \implies B_1 \preccurlyeq_{\mathsf{bl}} B_2$.

THEOREM 2 (FULL ABSTRACTION OF $\sqsubseteq_{\mathsf{q}}$) $B_1 \preccurlyeq_{\mathsf{bl}} B_2 \implies B_1 \sqsubseteq_{\mathsf{q}} B_2$.

As a corollary of the above theorems, a program transformation $B_2 \rightsquigarrow B_1$ is valid if and only if $B_1 \sqsubseteq_{\mathsf{q}} B_2$ holds. We prove Theorem 1 in §B. We give a proof sketch of Theorem 2 in §8 and a full proof in §F.

### 4.4 Example Transformation

We now consider how our approach applies to a simple program transformation:

$B_2$: `store(x,l1); store(x,l1)` $\rightsquigarrow$ $B_1$: `store(x,l1)`

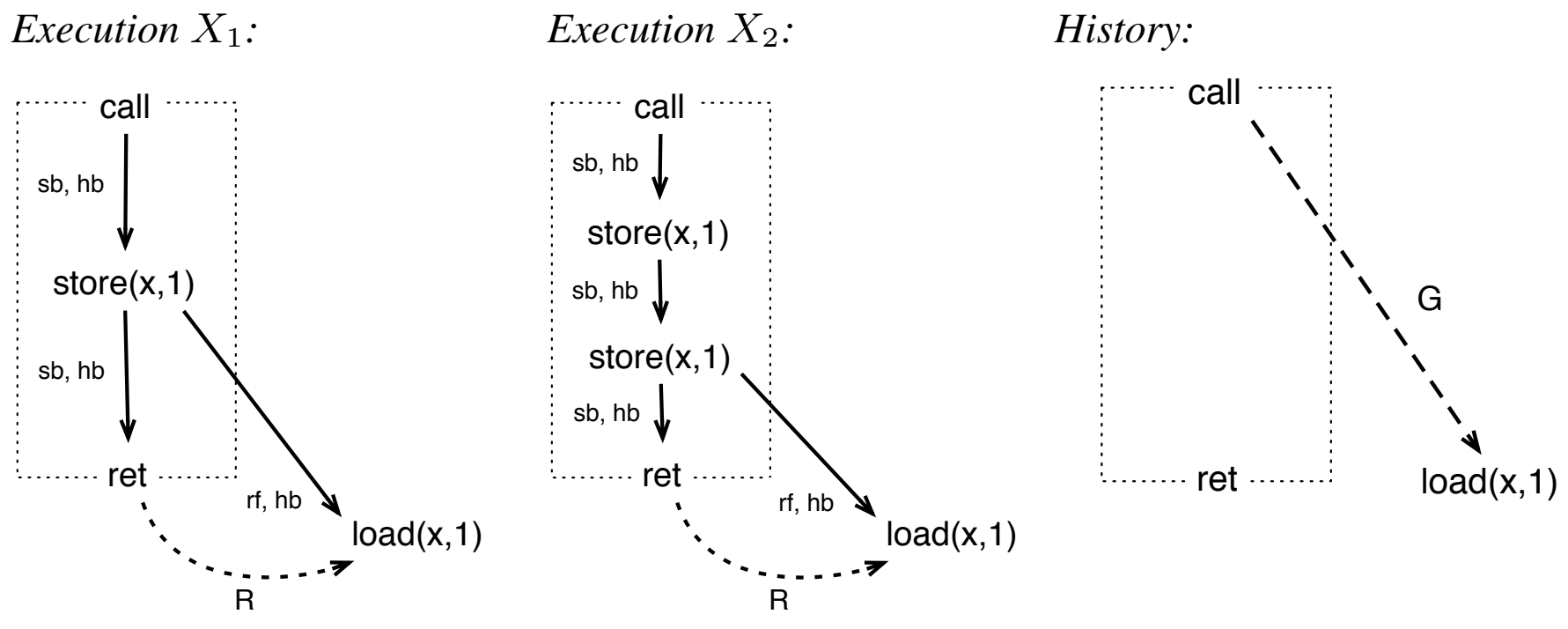

**Fig. 6.** History comparison for an example program transformation.

To verify this transformation, we must show that $B_1 \sqsubseteq_{\mathsf{q}} B_2$. To do this, we must consider the unboundedly many block-local executions. Here we just illustrate the reasoning for a single block-local execution; in §5 below we define a context reduction which lets us consider a finite set of such executions.

In Figure 6, we illustrate the necessary reasoning for an execution $X_1 \in [\![B_1, \mathcal{A}, R, S]\!]$, with a context action set $\mathcal{A}$ consisting of a single load $\mathtt{x} = 1$, a context relation $R$ relating ret to the load, and an empty $S$ relation. This choice of $R$ forces the context load to read from the store in the block. We can exhibit an execution $X_2 \in [\![B_2, \mathcal{A}, R, S]\!]$ with a matching history by making the context load read from the final store in the block.

## 5 A Finite Denotation

The approach above simplifies contexts by removing syntax and non-hb structure, but there are still infinitely many $\mathcal{A}/R/S$ contexts for any code-block. To solve this, we introduce a type of context reduction which allows us to consider only finitely many block-local executions. This means that we can automatically check transformations by examining all such executions. However this 'cut down' approach is no longer fully abstract. We modify our denotation as follows:

– We remove the quantification over context relation $R$ from definition (7) by fixing it as $\emptyset$. In exchange, we extend the history with an extra component called a *deny*.
– We eliminate redundant block-local executions from the denotation, and only consider a reduced set of executions $X$ that satisfy a predicate $\mathsf{cut}(X)$.

These two steps are both necessary to achieve finiteness. Removing the $R$ relation reduces the amount of structure in the context. This makes it possible to then remove redundant patterns – for example, duplicate reads from the same write.

Before defining the two steps in detail, we give the structure of our modified refinement $\sqsubseteq_{\mathsf{c}}$. In the definition, $\mathsf{hist}_{\mathsf{E}}(X)$ stands for the *extended history* of an execution $X$, and $\sqsubseteq_{\mathsf{E}}$ for refinement on extended histories.

$$B_1 \sqsubseteq_{\mathsf{c}} B_2 \iff^{\Delta} \forall \mathcal{A}, S. \forall X_1 \in [\![B_1, \mathcal{A}, \emptyset, S]\!]. \\ \mathsf{cut}(X_1) \implies \exists X_2 \in [\![B_2, \mathcal{A}, \emptyset, S]\!]. \mathsf{hist}_{\mathsf{E}}(X_1) \sqsubseteq_{\mathsf{E}} \mathsf{hist}_{\mathsf{E}}(X_2) \quad (8)$$

As with $\sqsubseteq_{\mathsf{q}}$ above, the refinement $\sqsubseteq_{\mathsf{c}}$ is adequate. However, it is not fully abstract (we provide a counterexample in §D). We prove the following theorem in §E.

THEOREM 3 (ADEQUACY OF $\sqsubseteq_{\mathsf{c}}$) $B_1 \sqsubseteq_{\mathsf{c}} B_2 \implies B_1 \preccurlyeq_{\mathsf{bl}} B_2$.

### 5.1 Cutting Predicate

Removing the context relation $R$ in definition (8) removes a large amount of structure from the context. However, there are still unboundedly many block-local executions with an empty $R$ – for example, we can have an unbounded number of reads and writes that do not interact with the block. The cutting predicate identifies these redundant executions.

We first identify the actions in a block-local execution that are *visible*, meaning they directly interact with the block. We write $\mathsf{code}(X)$ for the set of actions in $X$ generated by the code-block. Visible actions belong to $\mathsf{code}(X)$, read from $\mathsf{code}(X)$, or are read by $\mathsf{code}(X)$. In other words,

$$\mathsf{vis}(X) \triangleq \mathsf{code}(X) \cup \{u \mid \exists v \in \mathsf{code}(X).\, u \xrightarrow{\mathsf{rf}} v \vee v \xrightarrow{\mathsf{rf}} u\}$$

Informally, cutting eliminates three redundant patterns: *(i)* non-visible context reads, i.e. reads from context writes; *(ii)* duplicate context reads from the same write; and *(iii)* duplicate non-visible writes that are not separated in mo by a visible write. Formally we define $\mathsf{cut}'(X)$, the conjunction of cutR for read, and cutW for write.

$$\begin{aligned} \mathsf{cutR}(X) &\iff^{\Delta} \mathsf{reads}(X) \subseteq \mathsf{vis}(X) \wedge \\ &\qquad \forall r_1, r_2 \in \mathsf{contx}(X).\, (r_1 \neq r_2 \Rightarrow \neg\exists w.\, w \xrightarrow{\mathsf{rf}} r_1 \wedge w \xrightarrow{\mathsf{rf}} r_2) \\ \mathsf{cutW}(X) &\iff^{\Delta} \forall w_1, w_2 \in (\mathsf{contx}(X) \setminus \mathsf{vis}(X)). \\ &\qquad w_1 \xrightarrow{\mathsf{mo}} w_2 \Rightarrow \exists w_3 \in \mathsf{vis}(X).\, w_1 \xrightarrow{\mathsf{mo}} w_3 \xrightarrow{\mathsf{mo}} w_2 \\ \mathsf{cut}'(X) &\iff^{\Delta} \mathsf{cutR}(X) \wedge \mathsf{cutW}(X) \end{aligned}$$

The final predicate $\mathsf{cut}(X)$ extends this in order to keep LL-SC pairs together: it requires that, if $\mathsf{cut}'()$ permits one half of an LL-SC, the other is also permitted implicitly (for brevity we omit the formal definition of $\mathsf{cut}()$ in terms of $\mathsf{cut}'$).

It should be intuitively clear why the first two of the above patterns are redundant. The main surprise is the third pattern, which preserves some non-visible writes. This is required by Theorem 3 for technical reasons connected to per-location coherence. We illustrate the application of $\mathsf{cut}()$ to a block-local execution in Figure 7.

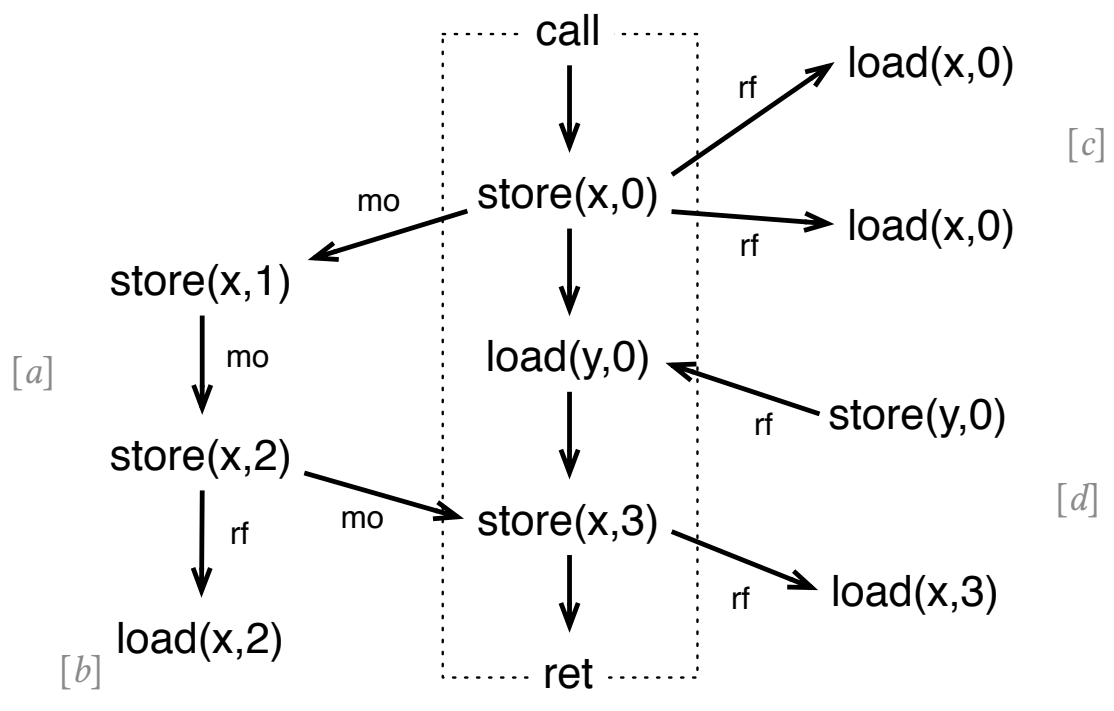

[a] *Forbidden by* cutW(). Two non-visible stores without a visible store intervening in mo.

[b] *Forbidden by* cutR(). Load is non-visible as it reads from a context store.

[c] *Forbidden by* cutR(). Both reads are visible, but are duplicates, reading from the same write.

[d] *Allowed.* Visible load and store.

**Fig. 7.** *Left:* block-local execution which includes patterns forbidden by cut(). *Right:* key explaining the patterns forbidden or allowed.

### 5.2 Extended History ($\mathsf{hist_E}$)

In our approach, each block-local execution represents a pattern of interaction between block and context. In our previous definition of $\sqsubseteq_\mathsf{q}$, constraints imposed by the block are captured by the guarantee, while constraints imposed by the context are captured by the $R$ relation. The definition (8) of $\sqsubseteq_\mathsf{c}$ removes the context relation $R$, but these constraints must still be represented. Instead, we replace $R$ with a history component called a *deny*. This simplifies the block-local executions, but compensates by recording more in the denotation.

The deny records the hb edges that *cannot* be enforced due to the execution structure. For example, consider the block-local execution[6] of Figure 8.

This pattern could not occur in a context that generates the dashed edge $D$ as a hb – to do so would violate the HBvsMO axiom. In our previous definition of $\sqsubseteq_\mathsf{q}$, we explicitly represented the presence or absence of this edge through the $R$ relation. In our new formulation, we represent such 'forbidden' edges in the history by a deny edge.

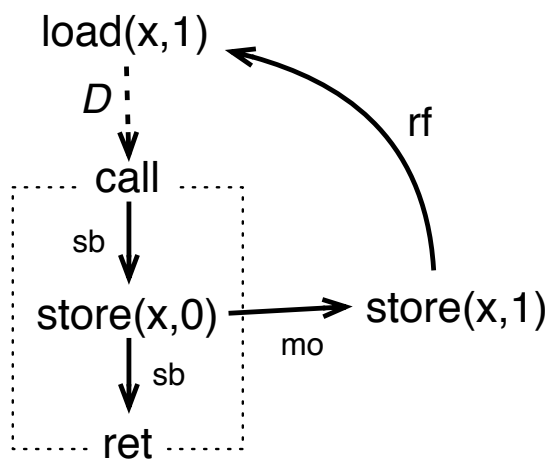

**Fig. 8.** A deny edge.

The *extended history* of an execution $X$, written $\mathsf{hist_E}(X)$ is a triple $(\mathcal{A}, G, D)$, consisting of the familiar notions of action set $\mathcal{A}$ and guarantee $G \subseteq \mathcal{A} \times \mathcal{A}$, together with deny $D \subseteq \mathcal{A} \times \mathcal{A}$ as defined below:

$$D \triangleq \{(u,v) \mid \mathsf{HBvsMO\text{-}d}(u,v) \vee \mathsf{Cohere\text{-}d}(u,v) \vee \mathsf{RFval\text{-}d}(u,v)\} \ \cap \\ \big((\mathsf{contx}(X) \times \mathsf{contx}(X)) \cup (\mathsf{contx}(X) \times \{\mathsf{call}\}) \cup (\{\mathsf{ret}\} \times \mathsf{contx}(X))\big)$$

Each of the predicates HBvsMO-d, Cohere-d, and RFval-d generates the deny for one validity axiom. In the diagrammatic definitions below, dashed edges represent the deny edge, and $\mathsf{hb}^*$ is the reflexive-transitive closure of hb:

[6] We use this execution for illustration, but in fact the cut() predicate would forbid the load.

$$\mathsf{HBvsMO\text{-}d}(u,v):\ \ \exists w_1, w_2.\ w_1 \xrightarrow{\mathsf{hb}^*} u \overset{D}{\dashrightarrow} v \xrightarrow{\mathsf{hb}^*} w_2 \quad \text{with } w_2 \xrightarrow{\mathsf{mo}} w_1$$

$$\mathsf{Coherence\text{-}d}(u,v):\ \ w_1 \xrightarrow{\mathsf{mo}} w_2 \xrightarrow{\mathsf{hb}^*} u \overset{D}{\dashrightarrow} v \xrightarrow{\mathsf{hb}^*} r \quad \text{with } w_1 \xrightarrow{\mathsf{rf}} r$$

$$\mathsf{RFval\text{-}d}(u,v):\ \ \exists w, r.\, \mathsf{gvar}(w) = \mathsf{gvar}(r) \land \neg\exists w'.\, w' \xrightarrow{\mathsf{rf}} r \land\ w \xrightarrow{\mathsf{hb}^*} u \overset{D}{\dashrightarrow} v \xrightarrow{\mathsf{hb}^*} r$$

One can think of a deny edge as an 'almost' violation of an axiom. For example, if $\mathsf{HBvsMO\text{-}d}(u,v)$ holds, then the context cannot generate an extra hb-edge $u \xrightarrow{\mathsf{hb}} v$ – to do so would violate HBvsMO.

Because deny edges represent constraints on the context, weakening the deny places fewer constraints, allowing more behaviours, so we compare them with relational inclusion:

$$(\mathcal{A}_2, G_2, D_2) \sqsubseteq_{\mathsf{E}} (\mathcal{A}_2, G_2, D_2) \ \overset{\Delta}{\iff}\ \mathcal{A}_1 = \mathcal{A}_2 \land G_2 \subseteq G_1 \land D_2 \subseteq D_1$$

This refinement on extended histories is used to define our refinement relation on blocks, $\sqsubseteq_{\mathsf{c}}$, def. (8).

### 5.3 Finiteness

THEOREM 4 (FINITENESS) *If for a block $B$ and state $\sigma$ the set of thread-local executions $\langle B, \sigma\rangle$ is finite, then so is the set of resulting block-local executions, $\{X \mid \exists \mathcal{A}, S.\, X \in [\![B, \mathcal{A}, \emptyset, S]\!] \land \mathsf{cut}(X)\}$.*

*Proof (sketch).* It is easy to see for a given thread-local execution there are finitely many possible visible reads and writes. Any two non-visible writes must be distinguished by at least one visible write, limiting their number. □

Theorem 4 means that any transformation can be checked automatically if the two blocks have finite sets of thread-local executions. We assume a finite data domain, meaning action can only take finitely many distinct values in Val. Recall also that our language does not include loops. Given these facts, any transformations written in our language will satisfy finiteness, and can therefore by automatically checked.

## 6 Prototype Verification Tool

Stellite is our prototype tool that verifies transformations using the Alloy* model checker [11, 17]. Our tool takes an input transformation $B_2 \rightsquigarrow B_1$ written in a C-like syntax. It automatically converts the transformation into an Alloy* model encoding $B_1 \sqsubseteq_{\mathsf{c}} B_2$. If the tool reports success, then the transformation is verified for unboundedly large syntactic contexts and executions.

An Alloy model consists of a collection of predicates on relations, and an instance of the model is a set of relations that satisfy the predicates. As previously noted in [28], there is therefore a natural fit between Alloy models and axiomatic memory models.

At a high level, our tool works as follows:

1. The two sides of an input transformation $B_1$ and $B_2$ are automatically converted into Alloy predicates expressing their syntactic structure. Intuitively, these block predicates are built by following the thread-local semantics from §3.
2. The block predicates are linked with a pre-defined Alloy model expressing the memory model and $\sqsubseteq_{\mathsf{c}}$.
3. The Alloy* solver searches (using SAT) for a history of $B_1$ that has no matching history of $B_2$. We use the higher-order Alloy* solver of [17] because the standard Alloy solver cannot support the existential quantification on histories in $\sqsubseteq_{\mathsf{c}}$.

The Alloy* solver is parameterised by the maximum size of the model it will examine. However, our finiteness theorem for $\sqsubseteq_{\mathsf{c}}$ (Theorem 4) means there is a bound on the size of cut-down context that needs to be considered to verify any given transformation. If our tool reports that a transformation is correct, it is verified in all syntactic contexts of unbounded size.

Given a query $B_1 \sqsubseteq_{\mathsf{c}} B_2$, the required context bound grows in proportion to the number of internal actions on distinct locations in $B_1$. This is because our cutting predicate permits context actions if they interact with internal actions, either directly, or by interleaving between internal actions. In our experiments we run the tool with a model bound of 10, sufficient to give soundness for all the transformations we consider. Note that most of our example transformations do not require such a large bound, and execution times improve if it is reduced.

If a counter-example is discovered, the problematic execution and history can be viewed using the Alloy model visualiser, which has a similar appearance to the execution diagrams in this paper. The output model generated by our tool encodes the history of $B_1$ for which no history of $B_2$ could be found. As $\sqsubseteq_{\mathsf{c}}$ is not fully abstract, this counter-example could, of course, be spurious.

Stellite currently supports transformations on code-blocks with atomic reads, writes, and fences. It does not yet support code-blocks with non-atomic accesses (see §7), LL-SC, or branching control-flow. We believe supporting the above features would not present fundamental difficulties, since the structure of the Alloy encoding would be similar. Despite the above limitations, our prototype demonstrates that our cut-down denotation can be used for automatic verification of important program transformations.

*Experimental results.* We have tested our tool on a range of different transformations. A table of experimental results is given in Figure 9. Many of our examples are derived from [23] – we cover all their examples that fit into our tool's input language. Transformations of the sort that we check have led to real-world bugs in GCC [18] and LLVM [8]. Note that some transformations are invalid because of their effect on local variables, e.g. $\texttt{skip} \rightsquigarrow l := \texttt{load}(x)$. The closely related transformation $\texttt{skip} \rightsquigarrow \texttt{load}(x)$ throws away the result of the read, and is consequently valid.

Our tool takes significant time to verify some of the above examples, and two of the transformations cause the tool to time out. This is due to the complexity and non-determinism of the C11 model. In particular, our execution times are comparable to existing C++ model *simulators* such as Cppmem when they run on a few lines of code [3]. However, our tool is a sound transformation verifier, rather than a simulator, and thus solves a more difficult problem: transformations are verified for unboundedly large syntactic contexts and executions, rather than for a single execution.

| Introduction, validity, time (s) | | |
|---|---|---|
| $\texttt{skip} \rightsquigarrow \texttt{fc}$ | ✓ | 76 |
| $\texttt{skip} \rightsquigarrow \texttt{ld}(x)$ | ✓ | 429 |
| $\texttt{skip} \rightsquigarrow l := \texttt{ld}(x)$ | x | 18 |
| $l := \texttt{ld}(x) \rightsquigarrow l := \texttt{ld}(x); \texttt{st}(x,l)$ | x | 72 |
| $l := \texttt{ld}(x) \rightsquigarrow l := \texttt{ld}(y); l := \texttt{ld}(x)$ | ? | $\infty$ |
| $l := \texttt{ld}(x) \rightsquigarrow l := \texttt{ld}(x); l := \texttt{ld}(x)$ | ✓ | 20**k** |
| $\texttt{st}(x,l) \rightsquigarrow \texttt{st}(x,l); \texttt{st}(x,l)$ | x | 136 |
| $\texttt{fc} \rightsquigarrow \texttt{fc}; \texttt{fc}$ | ✓ | 248 |

| Elimination, validity, time (s) | | |
|---|---|---|
| $\texttt{fc} \rightsquigarrow \texttt{skip}$ | x | 15 |
| $l := \texttt{ld}(x) \rightsquigarrow \texttt{skip}$ | x | 17 |
| $l := \texttt{ld}(x); \texttt{st}(x,l) \rightsquigarrow l := \texttt{ld}(x)$ | x | 64 |
| $l := \texttt{ld}(x); l := \texttt{ld}(x) \rightsquigarrow l := \texttt{ld}(x)$ | ✓ | 2**k** |
| $\texttt{st}(x,l); l := \texttt{ld}(x) \rightsquigarrow \texttt{st}(x,l)$ | ✓ | 9**k** |
| $\texttt{st}(x,m); \texttt{st}(x,l) \rightsquigarrow \texttt{st}(x,l)$ | ✓ | 24**k** |
| $\texttt{fc}; \texttt{fc} \rightsquigarrow \texttt{fc}$ | ✓ | 382 |

| Exchange, validity, time (s) | | |
|---|---|---|
| $\texttt{fc}; l := \texttt{ld}(x) \rightsquigarrow l := \texttt{ld}(x); \texttt{fc}$ | x | 26 |
| $\texttt{fc}; \texttt{st}(x,l) \rightsquigarrow \texttt{st}(x,l); \texttt{fc}$ | x | 50 |
| $l := \texttt{ld}(x); \texttt{fc} \rightsquigarrow \texttt{fc}; l := \texttt{ld}(x)$ | x | 79 |
| $\texttt{st}(x,l); \texttt{fc} \rightsquigarrow \texttt{fc}; \texttt{st}(x,l)$ | x | 145 |
| $l := \texttt{ld}(x); \texttt{st}(y,m) \rightsquigarrow \texttt{st}(y,m); l := \texttt{ld}(x)$ | x | 28 |
| $m := \texttt{ld}(y); l := \texttt{ld}(x) \rightsquigarrow l := \texttt{ld}(x); m := \texttt{ld}(y)$ | x | 118 |
| $\texttt{st}(y,m); l := \texttt{ld}(x) \rightsquigarrow l := \texttt{ld}(x); \texttt{st}(y,m)$ | ? | $\infty$ |
| $\texttt{st}(y,m); \texttt{st}(x,l) \rightsquigarrow \texttt{st}(x,l); \texttt{st}(y,m)$ | x | 641 |

**Fig. 9.** Results from executing Stellite on a 32 core 2.3GHz AMD Opteron, with 128GB RAM, over Linux 3.13.0-88 and Java 1.8.0_91. `load`/`store`/`fence` are abbreviated to `ld`/`st`/`fc`. ✓ and x denote whether the transformation satisfies $\sqsubseteq_{\mathsf{c}}$. $\infty$ denotes a timeout after 8 hours.

# 7 Transformations with Non-Atomics

We now extend our approach to *non-atomic* (i.e. unsynchronised) accesses. C11 non-atomics are intended to enable sequential compiler optimisations that would otherwise be unsound in a concurrent context. To achieve this, any concurrent read-write or write-write pair of non-atomic actions on the same location is declared a *data race*, which causes the whole program to have undefined behaviour. Therefore, adding non-atomics impacts not just the model, but also our denotation.

## 7.1 Memory Model with Non-atomics

Non-atomic loads and stores are added to the model by introducing new commands $\texttt{store}_{\mathsf{NA}}(x,l)$ and $l := \texttt{load}_{\mathsf{NA}}(x)$ and the corresponding kinds of actions: $\mathsf{store}_{\mathsf{NA}}, \mathsf{load}_{\mathsf{NA}} \in \mathsf{Kind}$. We let $\mathsf{NA}$ be the set of all actions of these kinds. We partition global variables so that they are either only accessed by non-atomics, or by atomics. We do not permit non-atomic LL-SC operations. Two new validity axioms ensure that non-atomics read from writes that happen before them, but not from stale writes:

- RFHBNA: $\forall w, r \in \mathsf{NA}.\, w \xrightarrow{\mathsf{rf}} r \implies w \xrightarrow{\mathsf{hb}} r$
- CoherNA: $\neg\exists w_1, w_2, r \in \mathsf{NA}.\; w_1 \xrightarrow{\mathsf{hb}} w_2 \xrightarrow{\mathsf{hb}} r$ with $w_1 \xrightarrow{\mathsf{rf}} r$

```
        store(y,0); storeNA(x,1);
storeNA(x,1); ‖ l1 := loadNA(x);
store(y,1);   ‖ l2 := load(y);
              ‖ l3 := loadNA(x);
```

```
        store(y,0); storeNA(x,1);
storeNA(x,1); ‖ l1 := loadNA(x);
store(y,1);   ‖ l3 := loadNA(x);
              ‖ l2 := load(y);
```

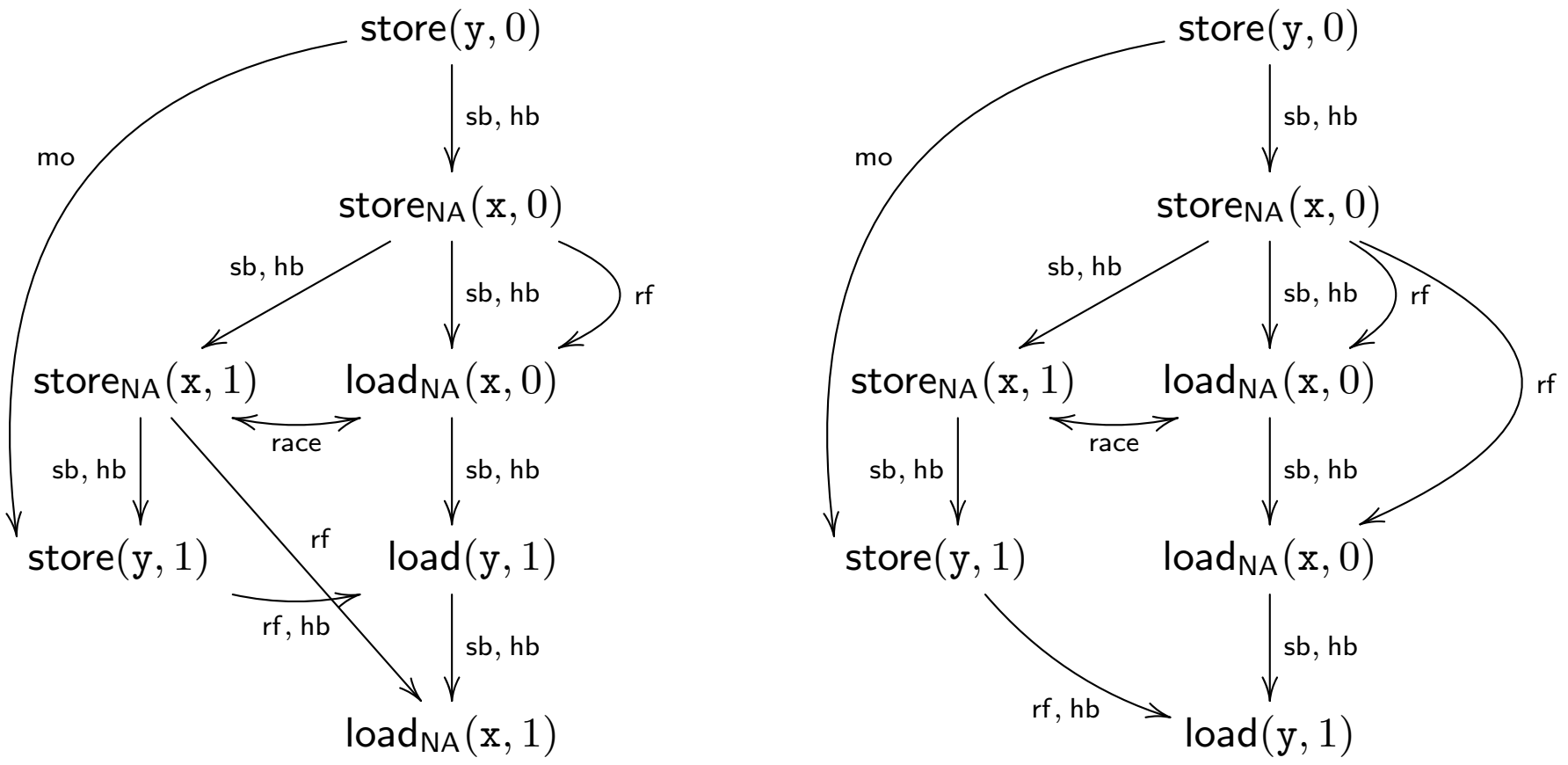

**Fig. 10.** *Top left:* augmented MP, with non-atomic accesses to x, and a new racy load. *Top right:* the same code optimised with $B_2 \rightsquigarrow B_1$. *Below each:* a valid execution.

Modification order (mo) does not cover non-atomic accesses, and we change the definition of happens-before (hb), so that non-atomic loads do not add edges to it:

– HBDEF: $\mathsf{hb} = (\mathsf{sb} \cup (\mathsf{rf} \cap \{(w, r) \mid w, r \notin \mathsf{NA}\}))^+$

Consider the code on the left in Figure 10: it is similar to MP from Figure 1, but we have removed the if-statement, made all accesses to x non-atomic, and we have added an additional load of x at the start of the right-hand thread. The valid execution of this code on the left-hand side demonstrates the additions to the model for non-atomics:

– modification order (mo) relates writes to atomic y, but not non-atomic x;
– the first load of x is forced to read from the initialisation by RFHBNA; and
– the second read of x is forced to read 1 because the hb created by the load of y obscures the now-stale initialisation write, in accordance with COHERNA.

The most significant change to the model is the introduction of a *safety axiom*, data-race freedom (DRF). This forbids non-atomic read-write and write-write pairs that are unordered in hb:

$$\text{DRF:} \quad \forall u, v \in \mathcal{A}. \begin{pmatrix} \exists x.\, u \neq v \wedge u = (\mathsf{store}(x, _)) \wedge \\ v \in \{(\mathsf{load}(x, _)), (\mathsf{store}(x, _))\} \end{pmatrix} \implies \begin{pmatrix} u \xrightarrow{\mathsf{hb}} v \vee v \xrightarrow{\mathsf{hb}} u \\ \vee\, u, v \notin \mathsf{NA} \end{pmatrix}$$

We write $\mathsf{safe}(X)$ if an execution satisfies this axiom. Returning to the left of Figure 10, we see that there is a violation of DRF – a race on non-atomics – between the first load of x and the store of x on the left-hand thread.

Let $[\![P]\!]_v^{\mathsf{NA}}$ be defined same way as $[\![P]\!]$ is in §3, def. (3), but with adding the axioms RFHBNA and COHERNA and substituting the changed axiom HBDEF. Then the semantics $[\![P]\!]$ of a program with non-atomics is:

$$[\![P]\!] \quad \stackrel{\Delta}{=} \quad \text{if } \forall X \in [\![P]\!]_v^{\mathsf{NA}}.\, \mathsf{safe}(X) \text{ then } [\![P]\!]_v^{\mathsf{NA}} \text{ else } \top$$

The undefined behaviour $\top$ subsumes all others, so any program observationally refines a racy program. Hence we modify our notion of observational refinement on whole programs:

$$P_1 \preccurlyeq_{\mathsf{pr}}^{\mathsf{NA}} P_2 \stackrel{\Delta}{\iff} (\mathsf{safe}(P_2) \implies (\mathsf{safe}(P_1) \wedge P_1 \preccurlyeq_{\mathsf{pr}} P_2))$$

This always holds when $P_2$ is unsafe; otherwise, it requires $P_1$ to preserve safety and observations to match. We define observational refinement on blocks, $\preccurlyeq_{\mathsf{bl}}^{\mathsf{NA}}$, by lifting $\preccurlyeq_{\mathsf{pr}}^{\mathsf{NA}}$ as per §2, def. (2).

### 7.2 Denotation with Non-atomics

We now define our denotation for non-atomics, $\sqsubseteq_{\mathsf{q}}^{\mathsf{NA}}$, building on the 'quantified' denotation $\sqsubseteq_{\mathsf{q}}$ defined in §4. (We have also defined a finite variant of this denotation using the cutting strategy described in §5 – we leave this to §C.)

Non-atomic actions do not participate in happens-before (hb) or coherence order (mo). For this reason, we need not change the structure of the history. However, non-atomics introduce undefined behaviour $\top$, which is a special kind of observable behaviour. If a block races with its context in some execution, the whole program becomes unsafe, for all executions. Therefore, our denotation must identify how a block may race with its context. In particular, for the denotation to be adequate, for any context $C$ and two blocks $B_1 \sqsubseteq_{\mathsf{q}}^{\mathsf{NA}} B_2$, we must have that if $C(B_1)$ is racy, then $C(B_2)$ is also racy.

To motivate the precise definition of $\sqsubseteq_{\mathsf{q}}^{\mathsf{NA}}$, we consider the following (sound) 'anti-roach-motel' transformation[7], noting that it might be applied to the right-hand thread of the code in the left of Figure 10:

$$\begin{aligned} &B_2 : \mathtt{l1} := \mathtt{load}_{\mathsf{NA}}(\mathtt{x});\ \mathtt{l2} := \mathtt{load}(\mathtt{y});\ \mathtt{l3} := \mathtt{load}_{\mathsf{NA}}(\mathtt{x}) \\ &\quad \leadsto \quad B_1 : \mathtt{l1} := \mathtt{load}_{\mathsf{NA}}(\mathtt{x});\ \mathtt{l3} := \mathtt{load}_{\mathsf{NA}}(\mathtt{x});\ \mathtt{l2} := \mathtt{load}(\mathtt{y}) \end{aligned}$$

In a standard roach-motel transformation [25], operations are moved into a synchronised block. This is sound because it only introduces new happens-before ordering between events, thereby restricting the execution of the program and preserving data-race freedom. In the above transformation, the second NA load of x is moved past the atomic load of y, effectively *out* of the synchronised block, reducing happens-before ordering, and possibly introducing new races. However, this is sound, because any data-race generated by $B_1$ must have already occurred with the first NA load of x, matching a racy execution of $B_2$. Verifying this transformation requires that we reason about races, so $\sqsubseteq_{\mathsf{q}}^{\mathsf{NA}}$ must account for both racy and non-racy behaviour.

[7] This example was provided to us by Lahav, Giannarakis and Vafeiadis in personal communication.

The code on the left of Figure 10 represents a context, composed with $B_2$, and the execution of Figure 10 demonstrates that together they are racy. If we were to apply our transformation to the fragment $B_2$ of the right-hand thread, then we would produce the code on the right in Figure 10. On the right in Figure 10, we present a similar execution to the one given on the left. The reordering on the right-hand thread has led to the second load of `x` taking the value 0 rather than 1, in accordance with RFHBNA. Note that the execution still has a race on the first load of `x`, albeit with different following events. As this example illustrates, when considering racy executions in the definition of $\sqsubseteq_{\mathsf{q}}^{\mathsf{NA}}$, we may need to match executions of the two code-blocks that behave differently after a race. This is the key subtlety in our definition of $\sqsubseteq_{\mathsf{q}}^{\mathsf{NA}}$.

In more detail, for two related blocks $B_1 \sqsubseteq_{\mathsf{q}}^{\mathsf{NA}} B_2$, if $B_2$ generates a race in a block-local execution under a given (reduced) context, then we require $B_1$ and $B_2$ to have corresponding histories *only up to the point the race occurs*. Once the race has occurred, the following behaviours of $B_1$ and $B_2$ may differ. This still ensures adequacy: when the blocks $B_1$ and $B_2$ are embedded into a syntactic context $C$, this ensures that a race can be reproduced in $C(B_2)$, and hence, $C(B_1) \preccurlyeq_{\mathsf{pr}}^{\mathsf{NA}} C(B_2)$.

By default, C11 executions represent a program's complete behaviour to termination. To allow us to compare executions up to the point a race occurs, we use *prefixes* of executions. We therefore introduce the *downclosure* $X^{\downarrow}$, the set of $(\mathsf{hb} \cup \mathsf{rf})^{+}$-prefixes of an execution $X$:

$$X^{\downarrow} \triangleq \{X' \mid \exists \mathcal{A}.\, X' = X|_{\mathcal{A}} \wedge \forall (u,v) \in (\mathsf{hb}(X) \cup \mathsf{rf}(X))^{+}.\, (v \in \mathcal{A} \Rightarrow u \in \mathcal{A})\}$$

Here $X|_{\mathcal{A}}$ is the projection of the execution $X$ to actions in $\mathcal{A}$. We lift the downclosure to sets of executions in the standard way.

Now we define our refinement relation $B_1 \sqsubseteq_{\mathsf{q}}^{\mathsf{NA}} B_2$ as follows:

$$\begin{aligned} B_1 \sqsubseteq_{\mathsf{q}}^{\mathsf{NA}} B_2 \overset{\Delta}{\iff} \;& \forall \mathcal{A}, R, S.\, \forall X_1 \in [\![B_1, \mathcal{A}, R, S]\!]_v^{\mathsf{NA}}.\, \exists X_2 \in [\![B_2, \mathcal{A}, R, S]\!]_v^{\mathsf{NA}}. \\ & (\mathsf{safe}(X_2) \implies \mathsf{safe}(X_1) \wedge \mathsf{hist}(X_1) \sqsubseteq_{\mathsf{h}} \mathsf{hist}(X_2)) \;\wedge \\ & (\neg\mathsf{safe}(X_2) \implies \exists X_2' \in (X_2)^{\downarrow}.\, \exists X_1' \in (X_1)^{\downarrow}. \\ & \qquad\qquad \neg\mathsf{safe}(X_2') \wedge \mathsf{hist}(X_1') \sqsubseteq_{\mathsf{h}} \mathsf{hist}(X_2')) \end{aligned}$$

In this definition, for each execution $X_1$ of block $B_1$, we witness an execution $X_2$ of block $B_2$ that is related. The relationship depends on whether $X_2$ is safe or unsafe.

- If $X_2$ is safe, then the situation corresponds to $\sqsubseteq_{\mathsf{q}}$ – see §4, def. (7). In fact, if $B_2$ is *certain* to be safe, for example because it has no non-atomic accesses, then the above definition is equivalent to $\sqsubseteq_{\mathsf{q}}$.
- If $X_2$ is unsafe then it has a race, and we do not have to relate the whole executions $X_1$ and $X_2$. We need only show that the race in $X_2$ is feasible by finding a prefix in $X_1$ that refines the prefix leading to the race in $X_2$. In other words, $X_2$ will behave consistently with $X_1$ *until it becomes unsafe*. This ensures that the race in $X_2$ will in fact occur, and its undefined behaviour will subsume the behaviour of $B_1$. After $X_2$ becomes unsafe, the two blocks can behave entirely differently, so we need not show that the complete histories of $X_1$ and $X_2$ are related.

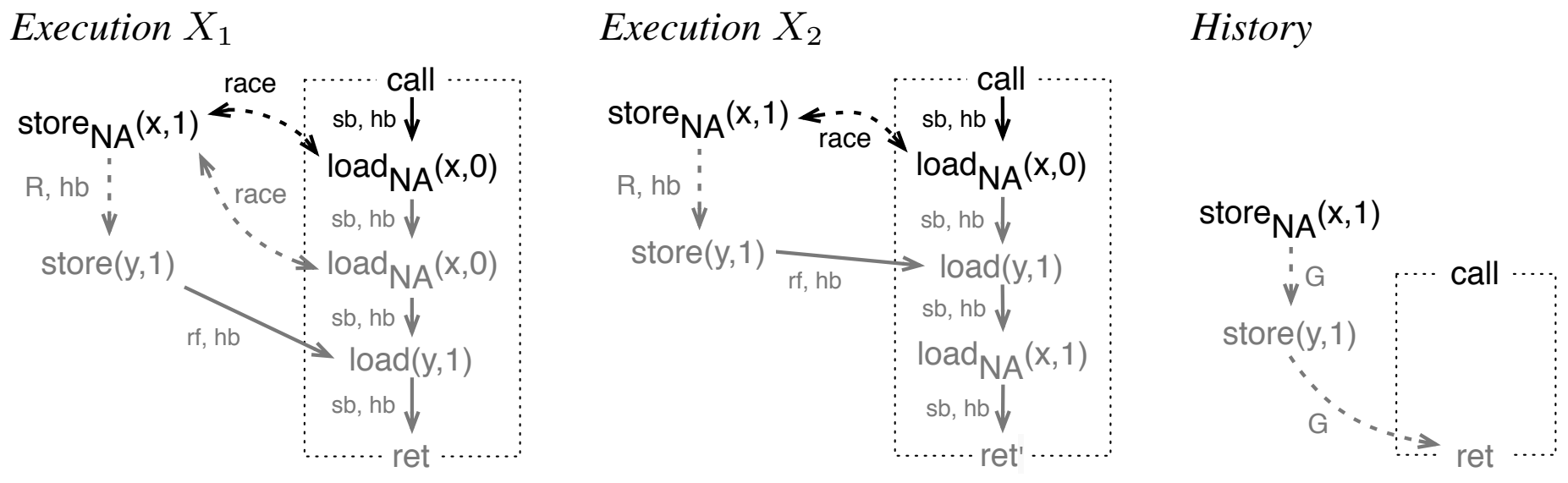

**Fig. 11.** History comparison for an NA-based program transformation

Recall the transformation $B_2 \rightsquigarrow B_1$ given above. To verify it, we must establish that $B_1 \sqsubseteq_{\mathsf{q}}^{\mathsf{NA}} B_2$. As before, we illustrate the reasoning for a single block-local execution – verifying the transformation would require a proof for all block-local executions.

In Figure 11 we give an execution $X_1 \in [\![B_1, \mathcal{A}, R, S]\!]$, with a context action set $\mathcal{A}$ consisting of a non-atomic store of $\mathtt{x} = 1$ and an atomic store of $\mathtt{y} = 1$, and a context relation $R$ relating the store of x to the store of y. Note that this choice of context actions matches the left-hand thread in the code listings of Figure 10, and there are data races between the loads and the store on x.

To prove the refinement for this execution, we exhibit a corresponding unsafe execution $X_2 \in [\![B_2, \mathcal{A}, R, S]\!]_v$. The histories of the *complete* executions $X_1$ and $X_2$ differ in their return action. In $X_2$ the load of y takes the value of the context store, so COHERNA forces the second load of x to read from the context store of x. This changes the values of local variables recorded in ret$'$. However, because $X_2$ is unsafe, we can select a prefix $X_2'$ which includes the race (we denote in grey the parts that we do not include). Similarly, we can select a prefix $X_1'$ of $X_1$. We have that $\mathsf{hist}(X_1') = \mathsf{hist}(X_2')$ (shown in the figure), even though the histories $\mathsf{hist}(X_1)$ and $\mathsf{hist}(X_2)$ do not correspond.

THEOREM 5 (ADEQUACY OF $\sqsubseteq_{\mathsf{q}}^{\mathsf{NA}}$) $B_1 \sqsubseteq_{\mathsf{q}}^{\mathsf{NA}} B_2 \implies B_1 \preccurlyeq_{\mathsf{bl}}^{\mathsf{NA}} B_2$.

THEOREM 6 (FULL ABSTRACTION OF $\sqsubseteq_{\mathsf{q}}^{\mathsf{NA}}$) $B_1 \preccurlyeq_{\mathsf{bl}}^{\mathsf{NA}} B_2 \Rightarrow B_1 \sqsubseteq_{\mathsf{q}}^{\mathsf{NA}} B_2$.

We prove Theorem 5 in §B and Theorem 6 in §F. Note that the prefixing in our definition of $\sqsubseteq_{\mathsf{q}}^{\mathsf{NA}}$ is required for full abstraction—but it would be adequate to always require *complete* executions with related histories.

## 8 Full Abstraction

The key idea of our proofs of full abstraction (Theorems 2 and 6, given in full in §F) is to construct a special syntactic context that is sensitive to one particular history. Namely, given an execution $X$ produced from a block $B$ with context happens-before $R$, this context $C_X$ guarantees: (1) that $X$ is the block portion of an execution of $C_X(B)$; and (2) for any block $B'$, if $C_X(B')$ has a different block history from $X$, then this is visible in different observable behaviour. Therefore for any blocks that are distinguished by different histories, $C_X$ can produce a program with different observable behaviour, establishing full abstraction.

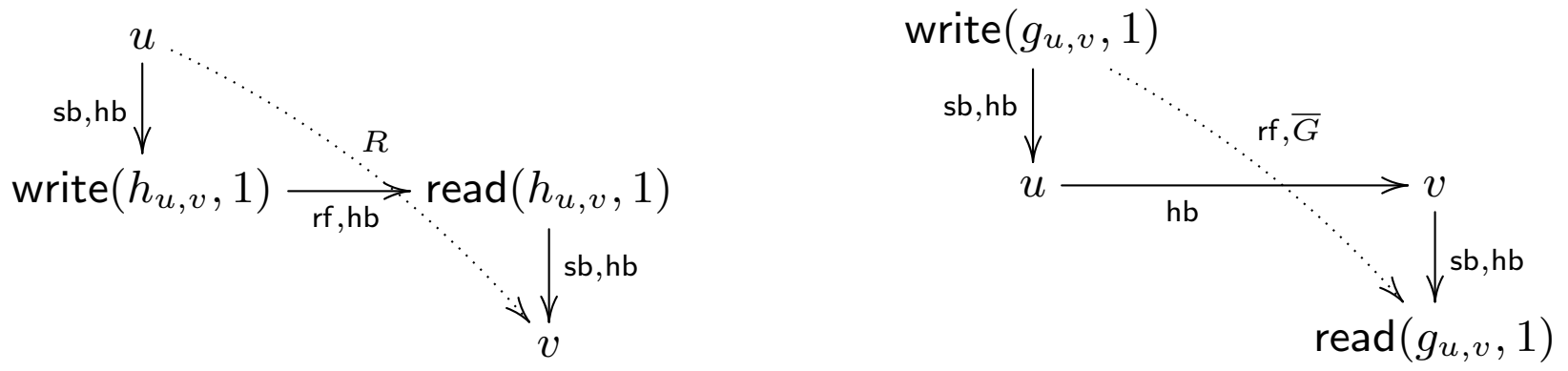

**Fig. 12.** The execution shapes generated by the special context for, on the *left*, generation of $R$, and on the *right*, errant history edges.

*Special context construction.* The precise definition of the special context construction $C_X$ is given in §F – here we sketch its behaviour. $C_X$ executes the context operations from $X$ in parallel with the block. It wraps these operations in auxiliary wrapper code to enforce context happens-before, $R$, and to check the history. If wrapper code fails, it writes to an error variable, which thereby alters the observable behaviour.

The context must generate edges in $R$. This is enforced by wrappers that use watchdog variables to create hb-edges: each edge $(u, v) \in R$ is replicated by a write and read on variable $h_{(u,v)}$. If the read on $h_{(u,v)}$ does not read the write, then the error variable is written. The shape of a successful read is given on the left in Figure 12.

The context must also prohibit history edges beyond those in the original guarantee $G$, and again it uses watchdog variables. For each $(u, v)$ *not* in $G$, the special context writes to watchdog variable $g_{(u,v)}$ before $u$ and a reads $g_{(u,v)}$ after $v$. If the read of $g_{(u,v)}$ *does* read the value written before $u$, then there is an errant history edge, and the error location is written. An erroneous execution has the shape given on the right in Figure 12 (omitting the write to the error location).

*Full abstraction and LL-SC.* Our proof of full abstraction for the language with C11 non-atomics requires the language to also include LL-SC, not just C11's standard CAS: the former operation increases the observational power of the context. However, *without* non-atomics (§4) CAS would be sufficient to prove full abstraction.

## 9 Related Work

Our approach builds on our prior work [3], which generalises linearizability [10] to the C11 memory model. This work represented interactions between a library and its clients by sets of histories consisting of a guarantee and a deny; we do the same for code-block and context. However, our previous work assumed *information hiding*, i.e., that the variables used by the library cannot be directly accessed by clients; we lift this assumption here. We also establish both adequacy and full abstraction, propose a finite denotation, and build an automated verification tool.

Our approach is similar in structure to the seminal concurrency semantics of Brookes [6]: i.e. a code block is represented by a denotation capturing possible interactions with an abstracted context. In [6], denotations are sets of traces, consisting of sequences of global program states; context actions are represented by changes in these states. To handle the more complex axiomatic memory model, our denotation consists of sets of context actions and relations on them, with context actions explicitly represented as such. Also, in order to achieve full abstraction, Brookes assumes a powerful atomic `await()` instruction which blocks until the global state satisfies a predicate. Our result does not require this: all our instructions operate on single locations, and our strongest instruction is LL-SC, which is commonly available on hardware.

Brookes-like approaches have been applied to several relaxed models: operational hardware models [7], TSO [12], and SC-DRF [20]. Also, [7, 20] define tools for verifying program transformations. All three approaches are based on traces rather than partial orders, and are therefore not directly portable to C11-style axiomatic memory models. All three also target substantially stronger (i.e. more restrictive) models.

Methods for verifying code transformations, either manually or using proof assistants, have been proposed for several relaxed models: TSO [24, 26, 27], Java [25] and C/C++ [23]. These methods are non-compositional in the sense that verifying a transformation requires considering the trace set of the entire program — there is no abstraction of the context. We abstract both the sequential and concurrent context and thereby support automated verification. The above methods also model transformations as rewrites on program executions, whereas we treat them directly as modifications of program syntax; the latter corresponds more closely to actual compilers. Finally, these methods all require considerable proof effort; we build an automated verification tool.

Our tool is a sound verification tool – that is, transformations are verified for all context and all executions of unbounded size. Several tools exist for testing (not verifying) program transformations on axiomatic memory models by searching for counter-examples to correctness, e.g., [15] for GCC and [8] for LLVM. Alloy was used by [28] in a testing tool for comparing memory models – this includes comparing language-level constructs with their compiled forms. Alloy has also been used in the MemSAT tool for simulation of the Java memory model [22]. Finally, our Alloy encoding is similar to the input files for the Herd/Cat memory model simulator [1].

## 10 Conclusions

We have proposed the first fully abstract denotational semantics for an axiomatic relaxed memory model, and using this, we have built the first tool capable of automatically verifying program transformation on such a model. Our theory lays the groundwork for further research into the properties of axiomatic models. In particular, our definition of the denotation as a set of histories and our context reduction should be portable to other axiomatic models based on happens-before, such as those for hardware [1].

**Acknowledgements.** Thanks to Jeremy Jacob, Viktor Vafeiadis, and John Wickerson for comments and suggestions. Dodds was supported by a Royal Society Industrial Fellowship, and undertook this work while faculty at the University of York. Batty is supported by a Lloyds Register Foundation and Royal Academy of Engineering Research Fellowship.

## A Collected Definitions

*The Thread-local semantics* of our target language. We write $\mathcal{A}_1 \uplus \mathcal{A}_2$ for a union that is defined only when actions in $\mathcal{A}_1$ and $\mathcal{A}_2$ use disjoint sets of identifiers. We omit identifiers from actions to avoid clutter.

$$\begin{aligned}
\langle l := \texttt{load}(x), \sigma\rangle &\triangleq \{(\{\mathsf{load}(x,a)\}, \emptyset, \sigma[l \mapsto a]) \mid a \in \mathsf{Val}\}\\
\langle \texttt{store}(x,l), \sigma\rangle &\triangleq \{(\{\mathsf{store}(x,a)\}, \emptyset, \sigma) \mid \sigma(l) = a\}\\
\langle l := \texttt{LL}(x), \sigma\rangle &\triangleq \{(\{\mathsf{LL}(x,a)\}, \emptyset, \sigma[l \mapsto a]) \mid a \in \mathsf{Val}\}\\
\langle l' := \texttt{SC}(x,l), \sigma\rangle &\triangleq \{(\{\mathsf{SC}(x,a)\}, \emptyset, \sigma[l' \mapsto 1]) \mid \sigma(l) = a\} \cup\\
&\qquad \{(\{\mathsf{SC}_f(x)\}, \emptyset, \sigma[l' \mapsto 0])\}\\
\langle \texttt{fence}, \sigma\rangle &\triangleq \{(\{\mathit{ll}, \mathit{sc}\}, \{(\mathit{ll}, \mathit{sc})\}, \sigma) \mid \mathit{ll} = \mathsf{LL}(\mathit{fen}, 0) \wedge \mathit{sc} = \mathsf{SC}(\mathit{fen}, 0)\}\\
\langle C_1 \parallel C_2, \sigma\rangle &\triangleq \{(\mathcal{A}_1 \uplus \mathcal{A}_2, \mathsf{sb}_1 \cup \mathsf{sb}_2, \sigma) \mid\\
&\qquad (\mathcal{A}_1, \mathsf{sb}_1, \sigma_1) \in \langle C_1, \sigma\rangle \wedge (\mathcal{A}_2, \mathsf{sb}_2, \sigma_2) \in \langle C_2, \sigma\rangle\}\\
\langle C_1; C_2, \sigma\rangle &\triangleq \{(\mathcal{A}_1 \uplus \mathcal{A}_2, \mathsf{sb}_1 \cup \mathsf{sb}_2 \cup (\mathcal{A}_1 \times \mathcal{A}_2), \sigma_2) \mid\\
&\qquad (\mathcal{A}_1, \mathsf{sb}_1, \sigma_1) \in \langle C_1, \sigma\rangle \wedge (\mathcal{A}_2, \mathsf{sb}_2, \sigma_2) \in \langle C_2, \sigma_1\rangle\}\\
\langle \mathsf{if}(l)\,\{C_1\}\,\mathsf{else}\,\{C_2\}, \sigma\rangle &\triangleq \begin{cases} \langle C_2, \sigma\rangle, & \text{if } \sigma(l) = 0\\ \langle C_1, \sigma\rangle, & \text{otherwise}\end{cases}
\end{aligned}$$

*Execution observational refinement*

$$X \preccurlyeq_{\mathsf{ex}} Y \quad \overset{\Delta}{\Longleftrightarrow} \quad \mathcal{A}(X|_{\mathsf{OVar}}) = \mathcal{A}(Y|_{\mathsf{OVar}}) \ \wedge\ \mathsf{hb}(Y|_{\mathsf{OVar}}) \subseteq \mathsf{hb}(X|_{\mathsf{OVar}})$$

*Program observational refinement*

$$P_1 \preccurlyeq_{\mathsf{pr}} P_2 \overset{\Delta}{\Longleftrightarrow} \forall X_1 \in [\![P_1]\!].\, \exists X_2 \in [\![P_2]\!].\, X_1 \preccurlyeq_{\mathsf{ex}} X_2$$

*Program observational refinement with NA*

$$P_1 \preccurlyeq^{\mathsf{NA}}_{\mathsf{pr}} P_2 \overset{\Delta}{\Longleftrightarrow} (\mathsf{safe}(P_2) \implies \mathsf{safe}(P_1) \wedge P_1 \preccurlyeq_{\mathsf{pr}} P_2)$$

*Block observational refinement*

$$B_1 \preccurlyeq_{\mathsf{bl}} B_2 \quad \overset{\Delta}{\Longleftrightarrow} \quad \forall C.\ C(B_1) \preccurlyeq_{\mathsf{pr}} C(B_2)$$

*History abstraction*

$$(\mathcal{A}_1, G_1) \sqsubseteq_{\mathsf{h}} (\mathcal{A}_2, G_2) \quad \overset{\Delta}{\Longleftrightarrow} \quad \mathcal{A}_1 = \mathcal{A}_2 \wedge G_2 \subseteq G_1$$

*Quantified abstraction*

$$\begin{aligned}
B_1 \sqsubseteq_{\mathsf{q}} B_2 \quad \overset{\Delta}{\Longleftrightarrow} \quad & \forall \mathcal{A}, R, S.\, \forall X_1 \in [\![B_1, \mathcal{A}, R, S]\!].\\
& \quad \exists X_2 \in [\![B_2, \mathcal{A}, R, S]\!].\, \mathsf{hist}(X_1) \sqsubseteq_{\mathsf{h}} \mathsf{hist}(X_2)
\end{aligned}$$

*Extended history abstraction*

$$(\mathcal{A}_2, G_2, D_2) \sqsubseteq_{\mathsf{E}} (\mathcal{A}_2, G_2, D_2) \overset{\Delta}{\iff} \mathcal{A}_1 = \mathcal{A}_2 \wedge G_2 \subseteq G_1 \wedge D_2 \subseteq D_1$$

*Cut abstraction*

$$B_1 \sqsubseteq_{\mathsf{c}} B_2 \overset{\Delta}{\iff} \forall \mathcal{A}, S. \forall X_1 \in [\![B_1, \mathcal{A}, \emptyset, S]\!]. \mathsf{cut}(X_1) \implies \exists X_2 \in [\![B_2, \mathcal{A}, \emptyset, S]\!]. \mathsf{hist}_{\mathsf{E}}(X_1) \sqsubseteq_{\mathsf{E}} \mathsf{hist}_{\mathsf{E}}(X_2)$$

*Cut predicates*

$$\mathsf{vis}(X) \overset{\Delta}{\iff} \mathsf{code}(X) \cup \{u \mid \exists v \in \mathsf{code}(X).\, u \xrightarrow{\mathsf{rf}} v \vee v \xrightarrow{\mathsf{rf}} u\}$$

$$\mathsf{cut}'(X) \overset{\Delta}{\iff} \mathsf{cutR}(X) \wedge \mathsf{cutW}(X)$$

$$\mathsf{cutR}(X) \overset{\Delta}{\iff} \mathsf{reads}(X) \subseteq \mathsf{vis}(X) \wedge \forall r_1, r_2 \in \mathsf{contx}(X).\, r_1 \neq r_2 \implies \neg \exists w.\, w \xrightarrow{\mathsf{rf}} r_1 \wedge w \xrightarrow{\mathsf{rf}} r_2$$

$$\mathsf{cutW}(X) \overset{\Delta}{\iff} \forall w_1, w_2 \in (\mathsf{contx}(X) \setminus \mathsf{vis}(X)).\, w_1 \xrightarrow{\mathsf{mo}} w_2 \implies \exists w_3 \in \mathsf{vis}(X).\, w_1 \xrightarrow{\mathsf{mo}} w_3 \xrightarrow{\mathsf{mo}} w_2$$

*Execution downclosure*

$$X^{\downarrow} \overset{\Delta}{=} \{X' \mid \exists \mathcal{A}.\, X' = X|_{\mathcal{A}} \wedge \forall (a, a') \in (\mathsf{hb}(X) \cup \mathsf{rf}(X))^{+}.\, a' \in \mathcal{A} \Rightarrow a \in \mathcal{A}\}$$

*Quantified abstraction with NA*

$$\begin{aligned} B_1 \sqsubseteq_{\mathsf{q}}^{\mathsf{NA}} B_2 \overset{\Delta}{\iff} & \forall \mathcal{A}, R, S. \forall X_1 \in [\![B_1, \mathcal{A}, R, S]\!]_v^{\mathsf{NA}}. \exists X_2 \in [\![B_2, \mathcal{A}, R, S]\!]_v^{\mathsf{NA}}. \\ & (\mathsf{safe}(X_2) \implies \mathsf{safe}(X_1) \wedge \mathsf{hist}(X_1) \sqsubseteq_{\mathsf{h}} \mathsf{hist}(X_2)) \wedge \\ & (\neg\mathsf{safe}(X_2) \implies \exists X_2' \in (X_2)^{\downarrow}. \exists X_1' \in (X_1)^{\downarrow}. \\ & \qquad \neg\mathsf{safe}(X_2') \wedge \mathsf{hist}(X_1') \sqsubseteq_{\mathsf{h}} \mathsf{hist}(X_2')) \end{aligned}$$

## B Proof of Theorems 1 and 5 (adequacy)

We now prove adequacy of $\sqsubseteq_{\mathsf{q}}^{\mathsf{NA}}$. As $\sqsubseteq_{\mathsf{q}}^{\mathsf{NA}} \implies \sqsubseteq_{\mathsf{q}}$, this suffices to prove adequacy of $\sqsubseteq_{\mathsf{q}}$. Our proof need several auxiliary notions:

- $\mathsf{codeE}(X)$ is the projection of an execution $X$ to actions in $(\mathsf{codeE}(X) \cup \mathsf{interf}(X) \cup \{\mathsf{call}, \mathsf{ret}\})$.
- The *interface actions* are actions on variables in $\mathsf{VS}_B$ that occur in the context. These are context actions that can affect the behaviour of the code-block. We write $\mathsf{interf}(X)$ for this set.
- $\mathsf{contxE}(X)$ is the projection of an execution $X$ to the context. This is a more complex projection than $\mathsf{codeE}(X)$ because it removes mo and rf over actions in $\mathsf{interf}(X)$. Let $\mathcal{I} = \mathsf{contx}(X) \cup \{\mathsf{call}, \mathsf{ret}\}$ and $\mathcal{C} = \mathsf{contx}(X) \setminus \mathsf{interf}(X)$. Then

$$\mathsf{contxE}(X) = (A(X)|_{\mathcal{I}}, \mathsf{hb}(X)|_{\mathcal{I}}, \mathsf{sb}(X)|_{\mathcal{I}}, \mathsf{mo}(X)|_{\mathcal{C}}, \mathsf{rf}(X)|_{\mathcal{C}})$$

- $\mathsf{hbC}(X)$ is the context-side projection of hb to interface actions. In other words, the projection of $\mathsf{hb}(X)$ to pairs in:

$$(\mathsf{interf}(X) \times \mathsf{interf}(X)) \cup (\mathsf{interf}(X) \times \{\mathsf{call}\}) \cup (\{\mathsf{ret}\} \times \mathsf{interf}(X))$$

- $\mathsf{atC}(X)$ is the context-side projection of at to context actions: i.e. the projection of $\mathsf{at}(X)$ to pairs in $(\mathsf{interf}(X) \times \mathsf{interf}(X))$.
- $[\![C, R, S]\!]_v$ is the *context-local* execution of a single-hole context $C$ – this is an analogous notion to the block-local execution, except that rf and mo are not generated for the interface. Here $R$ is a relation representing dependencies in hb arising from the code and $S$ represents code at edges. An execution $X$ is in this set iff:
  - $R$ is a code-side relation on interface actions $\mathsf{interf}(X)$:

$$R \subseteq (\mathsf{interf}(X) \times \mathsf{interf}(X)) \cup (\mathsf{interf}(X) \times \{\mathsf{ret}\}) \cup (\{\mathsf{call}\} \times \mathsf{interf}(X))$$

  - $S$ is a code-side relation on interface actions $\mathsf{interf}(X)$:

$$S \subseteq (\mathsf{interf}(X) \times \mathsf{interf}(X))$$

  - The execution satisfies the thread-local semantics:

$$(A(X), \mathsf{sb}(X)) \in \langle C \rangle$$

  We assume that a singleton hole has the following thread-local semantics:

$$\langle \{-\}, \sigma \rangle \quad \triangleq \quad \{(\{c, r\}, \{c \to r\}, \sigma') \mid c = \mathsf{call}(\sigma) \land r = \mathsf{ret}(\sigma')\}$$

  - $X$ satisfies HBDEF', ATOM$'$, ACYCLICITY, RFWF, HBVSMO, COHERENCE, RFHBNA, COHERNA.
  - The projection $X|_{\mathsf{contxE}(X)}$ satisfies RFVAL, MOWF. mo and rf are disjoint from actions in $\mathsf{interf}(X)$.

  We sometimes write $[\![C]\!]_v$ to stand for $[\![C, \emptyset, \emptyset]\!]$, i.e. the set of context-local executions with empty code-side relations.

LEMMA 7 (DECOMPOSITION) *Assume $X \in [\![C(B)]\!]_v$, and no there are no* at *edges in $C$ spanning $B$, nor any between the actions of $B$ and $C$. Then* $\mathsf{codeE}(X) \in [\![B, \mathsf{interf}(X), \mathsf{hbC}(X), \mathsf{atC}(X)]\!]_v$ *and* $\mathsf{contxE}(X) \in [\![C, \mathsf{hbL}(X), \mathsf{atL}(X)]\!]_v$.

*Proof (Proof (code)).* We have several proof obligations.

- $\mathsf{hbC}(X)$ and $\mathsf{atC}(X)$ are context-side relations on interface actions (trivial by definition).
- $(\mathsf{codeE}(\mathsf{codeE}(X)), \mathsf{sb}(\mathsf{codeE}(X))) \in \langle B \rangle$, i.e. the execution satisfies the thread-local semantics.
- The actions in $\mathsf{codeE}(\mathsf{codeE}(X))$ are in between a call / ret pair in sb. We assume we can introduce call / ret freely to satisfy this requirement.
- $\mathsf{codeE}(X)$ satisfies the validity axioms for a block-local execution – note that this replaces HBDEF with HBDEF', and ATOM with ATOM$'$.

For the first obligation, we argue inductively over the structure of $C$. First assume that $C = \{-\}$, i.e. $C$ consists only of a hole. In this case the result holds immediately from the thread-local semantics. For the inductive case, assume $C$ is a composite one-hole context, e.g. $C_1; C_2(-)$ / $C_1(-); C_2$ / $C_1 \| C_2(-)$ / etc.

For the fourth obligation, we prove $\mathsf{codeE}(X)$ satisfies the validity axioms by arguing in turn about each. Assume the following shorthand:

$$\mathsf{codeE}(X) = (A(l), \mathsf{hb}(l), \mathsf{at}(l)\mathsf{sb}(l), \mathsf{mo}(l), \mathsf{rf}(l))$$

HBDEF': Let $R = \mathsf{hbC}(X)$. Now prove in both directions:

$$(a,b) \in \mathsf{hb}(l) \implies (a,b) \in (\mathsf{sb}(l) \cup \mathsf{rf}_{\mathsf{AT}}(l) \cup R)^+ \quad (9)$$

$$(a,b) \in (\mathsf{sb}(l) \cup \mathsf{rf}_{\mathsf{AT}}(l) \cup R)^+ \implies (a,b) \in \mathsf{hb}(l) \quad (10)$$

For the first case, any $(a,b)$ in $\mathsf{hb}(l)$ must have code or interface actions at both ends, and must have originated from a path $(a,b) \in (\mathsf{sb}(X) \cup \mathsf{rf}_{\mathsf{AT}}(X))^+$. By construction, there are no rf-edges between $\mathsf{codeE}(X)$ and $\mathsf{contxE}(X)$. Therefore, portions of the path which stray into the context must enter and leave through call, ret, or actions in $\mathsf{interf}(X)$. These portions of the path will be summarised by $\mathsf{hbC}(X)$. As a result, for any such path, there must be an equivalent path $(a,b) \in (\mathsf{sb}(l) \cup \mathsf{rf}_{\mathsf{AT}}(l) \cup \mathsf{hbC}(X))^+$.

For the second case, we make a similar argument. For any pair $(c,d) \in \mathsf{hbC}(X)$, there must be a path $(c,d) \in (\mathsf{sb}(X) \cup \mathsf{rf}_{\mathsf{AT}}(X))$. As a consequence, for any $(a,b)$ in $(\mathsf{sb}(l) \cup \mathsf{rf}(l) \cup \mathsf{hbC}(X))^+$, there must be a path $(a,b) \in (\mathsf{sb}(X) \cup \mathsf{rf}_{\mathsf{NA}}(X))$. Thus $(a,b) \in \mathsf{hb}(X)$. As $\mathsf{hb}(l)$ is a projection of $\mathsf{hb}(X)$, this completes the proof.

ATOM′, ACYCLICITYRFWF, MOWF, COHERENCE, RFHBNA, COHERNA: all hold immediately by the fact that $\mathsf{codeE}(X)$ is a projection of $X$.

RFVAL: holds because $\mathsf{code}(X)$ contains exactly the actions in $X$ that are on locations $a \in \mathsf{gv}_B$. Therefore, the projection cannot remove the origin write for a read.

*Proof (Proof (context)).* Similar argument to the code.

LEMMA 8 (COMPLETION LEMMA) *Let $X$ be an execution. If* $\mathsf{valid}(X)$ *and* $(A(X), \mathsf{sb}(X)) \in \langle Q \rangle^{\downarrow}$, *then* $X \in [\![Q]\!]_v^{\downarrow}$.

*Proof.* We require the existence of a $Y \in [\![Q]\!]_v$ such that $X \in Y^{\downarrow}$. To prove this, we iteratively extend $X$ by adding sb-final actions, and show that the new execution can in each case be made valid. As all executions are finite, this proves the result.

Assume the current execution is $X_i$. We choose an $A(X_{i+1})$ and $\mathsf{sb}(X_{i+1})$ such that the new execution is extended by a single sb-final action, and that $(A(X_{i+1}), \mathsf{sb}(X_{i+1}) \in \langle Q \rangle^{\downarrow}$. We now need to show that we can construct a valid $X_{i+1}$.

Case-split on the type of the new action. Non-atomics read from their immediate hb predecessor, or the init value if none exists. Atomic reads read from the end of mo, and writes can be added to the end of mo. Compare-and-swaps read from the end of mo. All of these cases preserve the validity axioms.

Note that if the new action is a read, we may need to fix its value depending on an earlier write. This depends on the property of *receptiveness* – given a prefix $(A, \mathsf{sb}) \in \langle Q' \rangle$ and a read $r$ that is sb-maximal, any value can be given to the read. This property follows from the thread-local semantics: the only tricky cases are conditionals and LL-SC, where receptiveness is guaranteed by the fact that any possible value is represented in the set of possible reads.

LEMMA 9 (SAFETY COMPLETION) *Let $X, Y$ be valid executions.* $\neg\mathsf{safe}(X)$ *and* $X \in Y^{\downarrow}$ *implies* $\neg\mathsf{safe}(Y)$.

*Proof.* Prove the contrapositive: $\mathsf{safe}(Y) \implies \mathsf{safe}(X)$. This holds immediately from the fact that in a safe execution, potentially racy actions must be related in hb.

LEMMA 10 (COMPOSITION) *Let $X$ and $Y$ be executions such that* $X \in [\![B, \mathcal{A}, \mathsf{hbC}(Y), \mathsf{atC}(Y)]\!]_v^{\downarrow}$ *and* $Y \in [\![C, \mathcal{A}, R', S']\!]_v^{\downarrow}$ *with no* LL/SC *pairs crossing the block boundary in each case, with* $\mathsf{hist}(Y) \sqsubseteq_{\mathsf{h}} \mathsf{hist}(X)$ *and with* $\mathsf{atL}(X) = S'$. *Then there exists an execution $Z$ such that* $Z \in [\![C(B)]\!]_v^{\downarrow}$. *Furthermore:*

- *If* $\neg\mathsf{safe}(X)$ *or* $\neg\mathsf{safe}(Y)$, *then* $\neg\mathsf{safe}(Z)$.
- *If* $\mathsf{safe}(X)$, $\mathsf{safe}(Y)$, *and* $\mathsf{safe}(Z)$, *and* $X \in [\![B, \mathcal{A}, \mathsf{hbC}(Y), \mathsf{atC}(Y)]\!]_v$ *and* $Y \in [\![C, \mathcal{A}, R', S']\!]_v$, *then* $Z \in [\![C(B)]\!]_v$ *and* $\mathsf{contxE}(Y) \preccurlyeq_{\mathsf{ex}} \mathsf{contxE}(Z)$.

*Proof.* We begin by defining $Z$. Taking each term of the execution in turn:

- The action set $A(Z)$ is the union of the two action sets $A(X)$ and $A(Y)$, merging call, return and interface actions.
- $\mathsf{sb}(Z) = (\mathsf{sb}(X) \cup \mathsf{sb}(Y))^+$.
- $\mathsf{mo}_Z = (\mathsf{mo}(X) \cup \mathsf{mo}(Y))$ – as the two mo relations are disjoint, no transitive closure is needed.
- $\mathsf{rf}_Z = (\mathsf{rf}(X) \cup \mathsf{rf}(Y))$ – likewise.
- $\mathsf{hb}_Z = (\mathsf{sb}(Z) \cup \mathsf{rf}_{\mathsf{AT}}(Z))^+$, ie, according to HBDEF.
- $\mathsf{at}_Z = \mathsf{at}(X) \cup \mathsf{at}(Y)$.

We first need to show that $Z \in [\![C(L)]\!]_v^{\downarrow}$. To do this we use the completion lemma: thus our proof obligations are $(A(Z), \mathsf{sb}(Z)) \in \langle C(B_2) \rangle^{\downarrow}$ and $\mathsf{valid}(Z)$.

We observe that that $(A(Z), \mathsf{sb}(Z)) \in \langle C(B_2) \rangle^{\downarrow}$ is obvious from the thread-local semantics.

Next prove that $\mathsf{valid}(Z)$. HBDEF holds by construction. RFWF, RFVAL, MOWF, RBDEF are true trivially as for each variable, validity is checked solely in either the code or context. This leaves ACYCLICITY, HBVSMO, COHERENCE, COHERNA and ATOM. (RFHBNA needs to be treated specially – see below).

- For ACYCLICITY, a violation would correspond to a path in $(\mathsf{sb}(Z) \cup \mathsf{rf}_{\mathsf{AT}}(Z) \cup \mathsf{rf}_{\mathsf{NA}})^+$. As this path cannot appear in either $X$ or $Y$, it must cross between the two: each point where it does so must be an interface action or call / return. As a result, a corresponding violation can be constructed in $X$.

  Call-to-return paths are in $(\mathsf{sb}(X) \cup \mathsf{rf}_{\mathsf{AT}}(X))^+ \cup \mathsf{rf}_{\mathsf{NA}}(X))^+$. Conversely, return-to-call paths are in $(\mathsf{sb}(Y) \cup \mathsf{rf}_{\mathsf{AT}}(Y) \cup \mathsf{rf}_{\mathsf{NA}}(Y))^+$. As $Y$ satisfies RFHBNA, $\mathsf{rf}_{\mathsf{NA}}(Y) \in \mathsf{hb}(Y)$. Thus the return-to-call portions of the path are in $\mathsf{hbC}(Y)$. This contradicts the assumption that $X$ satisfies ACYCLICITY.

- For HBvsMO, a violation consists of a write pair $w_1, w_2$ such that $(w_1, w_2) \in \mathsf{hb}(Z)$ and $(w_2, w_1) \in \mathsf{mo}(Z)$. As mo is partitioned between code and context, either both writes are in $X$ or both in $Y$. By assumption, the violation is not solely in $X$ or $Y$, so the path from $w_1$ to $w_2$ in $(\mathsf{sb} \cup \mathsf{rf}_{\mathsf{AT}})^+$ must contain a sequence of interface actions or call / return.
  1. If the writes are in $X$, then mo is replicated immediately. The block-local portions of the path are in $(\mathsf{sb}(X) \cup \mathsf{rf}_{\mathsf{AT}}(X))^+$, while the context-local portions are in $\mathsf{hbC}(X)$. Thus we can replicate the violation.
  2. If the writes are in $Y$, we can use a similar argument. However, we also appeal to the fact that $\mathsf{hist}(Y) \sqsubseteq_{\mathsf{h}} \mathsf{hist}(X)$, which means that $\mathsf{hbL}(X) \subseteq \mathsf{hbL}(Y)$. This means that any code-side hb edge in $X$ can be replicated in $Y$ to recreate the violation.
- For COHERENCE and COHERNA, we note that rf and mo are partitioned between $X$ and $Y$. Therefore we can apply the same argument as for the previous axioms to show the hb edges for a violation must exist in either $X$ or $Y$.
- Similarly, for ATOM we note that at is partitioned between $X$ and $Y$ so any violation must exist in either $X$ or $Y$.

Finally, we consider RFHBNA. As $\mathsf{hist}Y \sqsubseteq_{\mathsf{h}} \mathsf{hist}X$, composing the two may weaken hb and generate violations on the context side. To solve this, we convert the RFHBNA violation to a safety violation. Take a $Z' \in Z^{\downarrow}$ such that there is a single $(\mathsf{hb} \cup \mathsf{rf})$-final RFHBNA violation. We redirect the origin of this read to its immediate hb-predecessor, or the initialisation value if this does not exist. This gives an execution $Z''$ which satisfies RFHBNA, but violated DRF. All the other validity axioms are preserved under prefixing, so by the completion lemma, $Z'' \in [\![C(B)]\!]_v^{\downarrow}$. We use $Z''$ as our constructed execution.

We now need to show that $\neg\mathsf{safe}(X)$ or $\neg\mathsf{safe}(Y)$ implies $\neg\mathsf{safe}(Z)$. If we had to fix an RFHBNA violation, the new execution $Z''$ is unsafe by construction. Otherwise, composition can only weaken hb, meaning any violation is trivially replicated.

Conversely, we need to show that if $\mathsf{safe}(X)$, $\mathsf{safe}(Y)$, and $\mathsf{safe}(Z)$, and $X \in [\![B, \mathcal{A}, \mathsf{hbC}(Y), \mathsf{atC}(Y)]\!]_v$ and $Y \in [\![C, \mathcal{A}, R', S']\!]_v$, then $Z \in [\![C(B)]\!]_v$ and $\mathsf{contxE}(Y) \preccurlyeq_{\mathsf{ex}} \mathsf{contxE}(Z)$. As $Z$ is safe, we know we did not have to fix a RFHBNA violation. For the rest of the proof, the same arguments as above give us a valid execution $Z \in [\![C(B)]\!]_v$.

It remains to show that $\mathsf{contxE}(Y) \preccurlyeq_{\mathsf{ex}} \mathsf{contxE}(Z)$. Inclusion of context actions follows from the construction of $Z$. Inclusion on hb follows from the fact that $\mathsf{hist}(Y) \sqsubseteq_{\mathsf{h}} \mathsf{hist}(X)$. Thus the composition can only weaken hb over context actions.

THEOREM 11 (ADEQUACY) *$B_1 \sqsubseteq_{\mathsf{q}}^{\mathsf{NA}} B_2 \implies B_1 \preccurlyeq_{\mathsf{bl}} B_2$ for blocks that include only matched* LL/SC *pairs.*

*Proof.* Our objective from the definition of $\preccurlyeq_{\mathsf{bl}}$ is the following property:

$$\begin{aligned}&\forall C, V.\ \neg\mathsf{safe}(C(B_2)) \lor \\ &\qquad (\mathsf{safe}(C(B_1)) \land \forall X \in [\![C(B_1)]\!]_v.\ \exists Y \in [\![C(B_2)]\!]_v.\ X|_V \preccurlyeq_{\mathsf{ex}} Y|_V)\end{aligned}$$

Begin the proof by picking an arbitrary $C, V$. The proof then proceeds by the normal steps: decomposition, abstraction, then composition.

- Case-split on whether $C(B_2)$ is safe or unsafe. If unsafe, we are done immediately. Therefore we can assume $\mathsf{safe}(C(B_2))$.
- Pick an arbitrary execution $X \in [\![C(B_1)]\!]_v$.
- Apply the decomposition lemma to show that that $\mathsf{contxE}(X) \in [\![C, \mathsf{hbL}(X), \mathsf{atL}(X)]\!]_v$ and $\mathsf{codeE}(X) \in [\![B_1, \mathsf{hbC}(X), \mathsf{atC}(X)]\!]_v$.
- Expand the definition of $\sqsubseteq_{\mathsf{q}}^{\mathsf{NA}}$, and pick $R = \mathsf{hbC}(X)$ and $S = \mathsf{atC}(X)$. This gives us executions $Y \in [\![B_2, \mathcal{A}, \mathsf{hbC}(X), \mathsf{atC}(X)]\!]_v^{\downarrow}$ and $X' \in \mathsf{codeE}(X)^{\downarrow}$ such that:

$$\begin{array}{l}\mathsf{hist}(X') \sqsubseteq_{\mathsf{h}} \mathsf{hist}(Y) \land \\ \quad \mathsf{safe}(Y) \implies (\mathsf{safe}(X') \land (X' = \mathsf{codeE}(X)) \land Y \in [\![B_2, \mathcal{A}, \mathsf{hbC}(X)]\!]_v)\end{array}$$

- Case-split on whether $\mathsf{safe}(Y) \land \mathsf{safe}(\mathsf{contxE}(X))$ holds. If not, then apply the composition lemma to build an execution $Z \in [\![C(B_2)]\!]_v^{\downarrow}$ such that $\neg\mathsf{safe}(Z)$. By lemma 9, there must exist a $Z' \in [\![C(B_2)]\!]_v$ such that $\neg\mathsf{safe}(Z')$, which contradicts our assumption that $C(B_2)$ is safe.
  Conversely, suppose $\mathsf{safe}(Y) \land \mathsf{safe}(\mathsf{contxE}(X))$ holds. In this case, we apply the context lemma to build a $Z \in [\![C(B_2)]\!]_v$ such that $\mathsf{contxE}(X) \preccurlyeq_{\mathsf{ex}} \mathsf{contxE}(Z)$. All actions on observable variables in $V$ must be be in the context, which means that $X|_V \preccurlyeq_{\mathsf{ex}} Z|_V$ must also hold.
  It remains to prove that $\mathsf{safe}(X)$ holds. First we observe that $\mathsf{safe}(\mathsf{codeE}(X))$ holds by the abstraction theorem. As $\mathsf{safe}(\mathsf{contxE}(X))$ also holds, the result follows immediately.

## C Non-atomics and Denies

We now define $\sqsubseteq_{\mathsf{c}}^{\mathsf{NA}}$, a refinement between denotations which includes both cutting and non-atomics.

To do this we first need extra deny shapes. In the following, the variables obey the following constraint:

$$u, a, c \in \mathsf{ret} \cup \mathsf{interf}(X) \qquad\qquad v, b, d \in \mathsf{call} \cup \mathsf{interf}(X)$$

All the actions $a, b, c, d, u, v$ are pairwise distinct. Note that some of the hb-edges are transitively closed, meaning that syntactically distinct actions might be the same – e.g. $w_1$ and $u$ in HBvsMO-d.

CoNA-d$(u, v)$:

$\exists w_1, w_2, r \in \mathsf{NA}.$

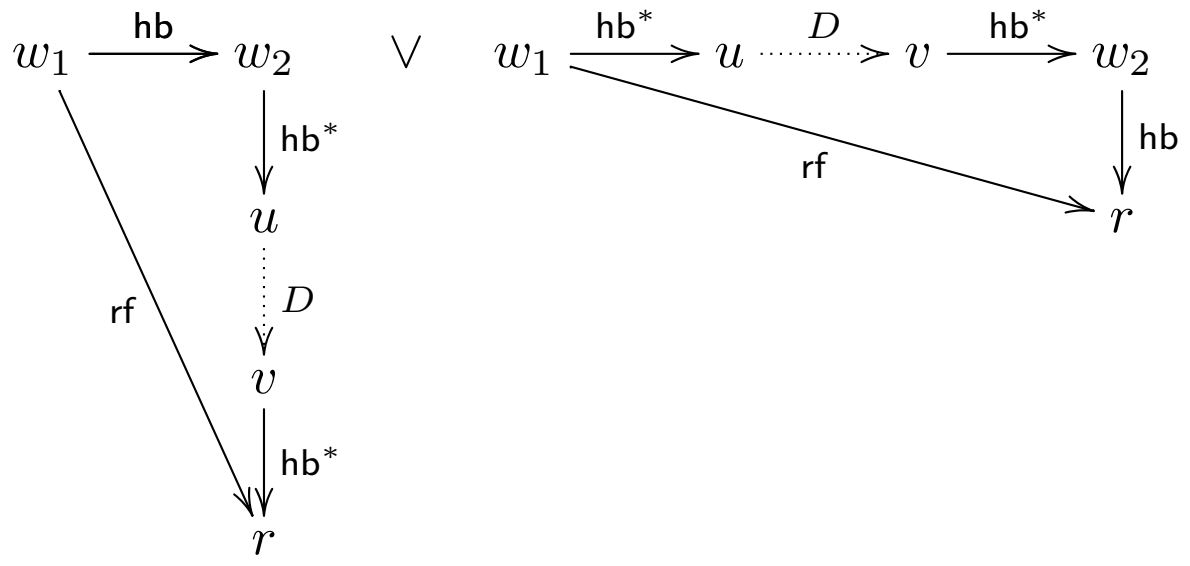

$\mathsf{CoNA\text{-}d2}(a,b,c,d)$: $\exists w_1, w_2, r \in \mathsf{NA}.$

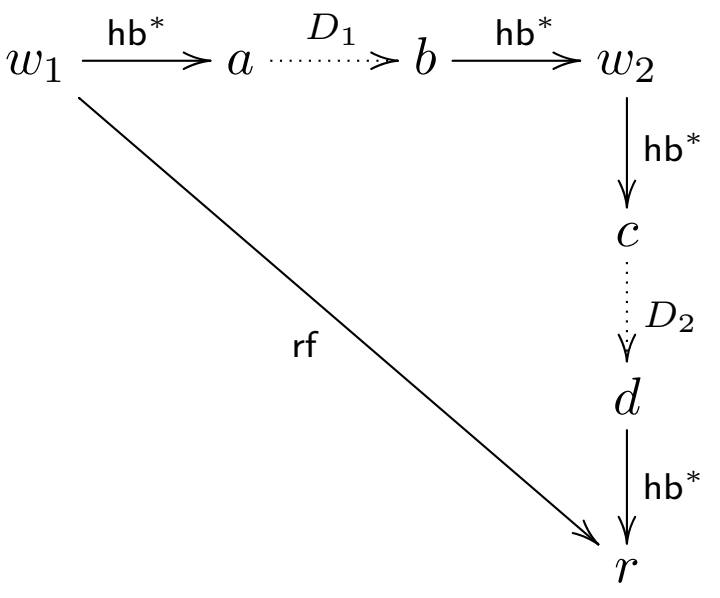

As before, we need a few notions to define the deny theorem.

- $\mathsf{denyL}(X)$ contains all the binary denies:

$$\mathsf{denyL}(X) \stackrel{\Delta}{=} \mathsf{HBvsMO\text{-}d} \cup \mathsf{CoWR\text{-}d} \cup \mathsf{Init\text{-}d} \cup \mathsf{CoNA\text{-}d}$$

- $\mathsf{denyNA}(X)$ contains the quaternary denies: $\mathsf{denyNA}(X) \stackrel{\Delta}{=} \mathsf{CoNA\text{-}d2}$
- $\mathsf{guarNA}(X)$ is the projection of $(\mathsf{rf}_{\mathsf{NA}} \cup \mathsf{hb})^+$ to pairs in

$$(\mathsf{interf}(X) \times \mathsf{interf}(X)) \cup (\mathsf{interf}(X) \times \{\mathsf{ret}\}) \cup (\{\mathsf{call}\} \times \mathsf{interf}(X))$$

- Let $\mathcal{I}$ be the set of actions $\mathsf{interf}(X) \cup \{\mathsf{call}, \mathsf{ret}\}$. The *augmented history* of $X$, written $\mathsf{hist}_{\mathsf{E}}(X)$, is defined as

$$\mathsf{hist}_{\mathsf{E}}(X) \stackrel{\Delta}{=} (A(X)|_{\mathcal{I}}, \mathsf{hbL}(X), \mathsf{denyL}(X), \mathsf{guarNA}(X), \mathsf{denyNA}(X))$$

- Two augmented histories, $H = (\mathcal{A}, G, D, M, N)$, $H' = (\mathcal{A}', G', D', M', N')$ are related $H \sqsubseteq_{\mathsf{h}} H'$ iff

$$\mathcal{A} = \mathcal{A}' \wedge G' \subseteq G \wedge D' \subseteq D \wedge M' \subseteq M \wedge N' \subseteq N$$

- $\mathsf{FinalNA}(X, a)$ holds if the action $a$ is (1) an NA action, and (2) is the $\mathsf{hb}$-final action in the code block in $X$.
- $\mathsf{hbA}(X, a)$, for an execution $X$ and action $a$ is the projection of $\mathsf{hb}(X)$ to pairs in $(\{a\} \times \mathsf{interf}(X)) \cup (\mathsf{interf}(X) \times \{a\})$
- The comparison $a \leq_{\mathsf{na}}^{X,Y} b$ ensures that $b$ participates in a race if $a$ does. Formally, the comparison holds if: (1) $a$ and $b$ are actions on the same location; (2) $b$ is a write if $a$ is a write; and (3) $\mathsf{hbA}(Y, b) \subseteq \mathsf{hbA}(X, a)$. The final condition is needed to ensure that history edges cannot spuriously prevent a race in $Y$.
- The set of *almost-valid* executions $[\![B, \mathcal{A}, R, S]\!]_{vr}$ is defined identically to the standard semantics, except that it permits RFHBNA not to hold. We write $[\![B, \mathcal{A}, S]\!]_{vr}$ stands for $[\![B, \mathcal{A}, \emptyset, S]\!]_{vr}$

We then define *deny abstraction* as follows:

$$\begin{aligned} B_1 \sqsubseteq_{\mathsf{d}}^{\mathsf{NA}} B_2 \stackrel{\Delta}{=}\ & \forall S.\, \forall X \in [\![B_1, \mathcal{A}, S]\!]_{vr}^{\downarrow}.\, \exists Y \in [\![B_2, \mathcal{A}, S]\!]_{vr}^{\downarrow}. \\ & \mathsf{hist}_{\mathsf{E}}(X) \sqsubseteq_{\mathsf{E}} \mathsf{hist}_{\mathsf{E}}(Y) \wedge \\ & (\forall a.\, \mathsf{FinalNA}(X, a) \implies \exists b \in A(Y).\, a \leq_{\mathsf{na}}^{X,Y} b) \wedge \\ & (X \in [\![B_1, \mathcal{A}, S]\!]_{vr} \implies Y \in [\![B_2, \mathcal{A}, S]\!]_{vr}) \end{aligned}$$

In addition to the cutting predicated defined in the body of the paper, we need the following to cover NA cuts.

$$\begin{aligned}\mathsf{NAcutR}(X) &\triangleq \forall r_1, r_2 \in (\mathsf{interf}(X) \cap \mathsf{Read} \cap \mathsf{NA}).\\ &\quad (\mathsf{val}(r_1) = \mathsf{val}(r_2) = \mathsf{init} \vee \exists w.\, w \xrightarrow{\mathsf{rf}} r_1 \wedge w \xrightarrow{\mathsf{rf}} r_2)\\ &\qquad \implies (r_1 = r_2)\\ \mathsf{NAcutW}(X) &\triangleq \forall w_1, w_2 \in (\mathsf{interf}(X) \cap \mathsf{Write} \cap \mathsf{NA}).\\ &\quad (\mathsf{loc}(w_1) = \mathsf{loc}(w_2)) \implies\\ &\qquad (w_1 = w_2) \vee (\exists r \in \mathsf{code}(X).\, w_1 \xrightarrow{\mathsf{rf}} r \vee w_2 \xrightarrow{\mathsf{rf}} r)\end{aligned}$$

The context cutting predicate is defined as the conjunction of these predicates:

$$\mathsf{cut}^{\mathsf{NA}}(X) \triangleq \mathsf{cutR}(X) \wedge \mathsf{cutW}(X) \wedge \mathsf{NAcutR}(X) \wedge \mathsf{NAcutW}(X)$$

We then define *cut abstraction* as follows:

$$\begin{aligned}B_1 \sqsubseteq_{\mathsf{c}}^{\mathsf{NA}} B_2 \;\triangleq\; & \forall X \in [\![B_1]\!]_{vr}^{\downarrow}.\, \mathsf{cut}^{\mathsf{NA}}(X) \implies\\ & \exists Y \in [\![B_2]\!]_{vr}^{\downarrow}.\, \mathsf{hist}_{\mathsf{E}}(X) \sqsubseteq_{\mathsf{E}} \mathsf{hist}_{\mathsf{E}}(Y) \wedge\\ & (\forall a.\, \mathsf{FinalNA}(X, a) \implies \exists b \in A(Y).\, a \leq_{\mathsf{na}}^{X,Y} b) \wedge\\ & (X \in [\![B_1]\!]_{vr} \implies Y \in [\![B_2]\!]_{vr})\end{aligned}$$

THEOREM 12 $B_1 \sqsubseteq_{\mathsf{d}}^{\mathsf{NA}} B_2 \implies B_1 \sqsubseteq_{\mathsf{q}}^{\mathsf{NA}} B_2$

*Proof.* Pick a context-side $\mathcal{A}$, $R$ and an execution $X \in [\![B_1, \mathcal{A}, R, S]\!]_v$. Case-split on $\mathsf{safe}(X)$ – suppose first that it does not hold.

- Pick a prefix $X' \in X$ and action $a \in A(X')$ such that (1) $X'$ contains precisely one safety violation, which includes $a$; and (2) $\mathsf{FinalNA}(X', a)$ holds.
- Generate a new execution $X''$ by building $\mathsf{hb}$ as $(\mathsf{sw} \cup \mathsf{sb})^+$ (i.e. kick out $R$). As all axioms but RFHBNA are preserved under reduction of $\mathsf{hb}$, $X' \in [\![B_1, \mathcal{A}, S]\!]_{vr}^{\downarrow}$.
- Apply the assumption to give an execution $Y' \in [\![B_2, \mathcal{A}, S]\!]_{vr}^{\downarrow}$, such that $\mathsf{hist}_{\mathsf{E}}(X'') \sqsubseteq_{\mathsf{E}} \mathsf{hist}_{\mathsf{E}}(Y')$. By the theorem, there must exist an action $b$ to the same location such that $a \leq_{\mathsf{na}} b$.
- Build $Y$ from $Y'$ by defining $\mathsf{hb}(Y)$ as $\mathsf{sb}(Y') \cup \mathsf{rf}_{\mathsf{NA}}(Y') \cup R$, and keeping other relations the same. We now need to establish that (1) $\mathsf{hist}(X') \sqsubseteq_{\mathsf{h}} \mathsf{hist}(Y)$; (2) $\neg\mathsf{safe}(Y)$; and (3) $\mathsf{valid}(Y)$.
- $\mathsf{hist}(X') \sqsubseteq_{\mathsf{h}} \mathsf{hist}(Y)$ holds from the fact that $\mathsf{hist}_{\mathsf{NA}}(X'') \sqsubseteq_{\mathsf{E}} \mathsf{hist}_{\mathsf{E}}(Y')$, and both $X'$ and $Y$ are derived by adding the same relation $R$.
- To show $\neg\mathsf{safe}(Y)$ we observe that action $a$ in $X'$ participates in a race. As actions in a code-block are sb-sequenced, the other action $c$ forming the race must be in $\mathsf{interf}(X')$. If $(b, c)$ does not form a race in $Y$, then $(b, c)$ or $(c, b)$ must be in $\mathsf{hb}(Y)$. Any such path must be in $R \cup \mathsf{hbL}(Y) \cup \mathsf{hbA}(Y', b)$. The corresponding path must exist in $R \cup \mathsf{hbL}(X) \cup \mathsf{hbA}(X'', b)$, which rules out the race in $X$ and contradicts the assumption.

- Finally, we need to prove that $\mathsf{valid}(Y)$. HBDEF' holds by construction. RFWF, RFVAL, MOWF, ATOM are invariant under adding hb-edges, and so follow immediately from $\mathsf{valid}(Y')$. This leaves ACYCLICITY, COHERENCE, HBVSMO, COHERNA, and RFHBNA.

  All but RFHBNA are covered by a deny (RFHBNA requires special treatment). A new violation of an axiom caused by edges from $R$ would induce a corresponding deny shape in $\mathsf{hist}_{\mathsf{E}}(Y')$. As $\mathsf{hist}_{\mathsf{E}}(X') \sqsubseteq_{\mathsf{E}} \mathsf{hist}_{\mathsf{E}}(Y)$ this deny shape must also be in $X'$. However, this means that the corresponding violation can be replicated in $X'$, which contradicts the assumption that $\mathsf{valid}(X')$ holds.
- Thus, we have an almost-valid execution $Y \in [\![B_2, \mathcal{A}, R]\!]^{\downarrow}_{vr}$ such that $\mathsf{hist}(X') \sqsubseteq_{\mathsf{h}} \mathsf{hist}(Y)$; (2) $\neg\mathsf{safe}(Y)$.

  To complete the proof, we need to fix violations of RFHBNA. We use the same approach as in the proof of Theorem 10: (1) build a shorter prefix in $Y^{\downarrow}$ which contains precisely one violation of RFHBNA; (2) redirect the read to a valid origin, using the receptiveness of the thread-local semantics. This redirection does alter the history, because non-atomic reads do not appear in the quantified history. This gives an execution $Y''$ such that (1) $Y'' \in [\![B_2, \mathcal{A}, R, S]\!]^{\downarrow}_{vr}$; (2) $\neg\mathsf{safe}(Y'')$; and (3) $\mathsf{hist}(Y'') \in \mathsf{hist}(Y^{\downarrow})$.

  We finally need to show that there exists an $X''' \in X^{\downarrow}$ such that $\mathsf{hist}(X''') \sqsubseteq_{\mathsf{h}} \mathsf{hist}(Y'')$. This necessarily exists by application of the history prefixing lemma. Note that $X'''$ may not necessarily be unsafe, but $Y''$ is guaranteed to be unsafe by construction.

Now suppose that $\mathsf{safe}(X)$ holds. We use essentially the same proof structure as above: the constructed $Y$ may be safe or unsafe, depending whether we need to fix violations of RFHBNA.

## D Counter-example to full abstraction

Finiteness has a cost: $\sqsubseteq_{\mathsf{c}}$ is not fully abstract. To see this, consider the optimisation $B_2$: $\mathtt{load}(x) \rightsquigarrow B_1$: $\mathtt{skip}$. It is easy to see that $B_1 \sqsubseteq_{\mathsf{q}} B_2$ holds: the new load can read from either a hb-earlier write action, or the initialisation if none exists. Neither case introduces an extra guarantee edge.

However, $B_1 \sqsubseteq_{\mathsf{c}} B_2$ does not hold. If the context contains a write $W$, then the load can either read from it or the initialisation. The former generates a hb-edge in the history, while the latter generates a deny from RFval-d – thus history inclusion does not hold.

## E Proof of Theorems 3 and 5 (cut soundness)

We now prove that $\sqsubseteq^{\mathsf{NA}}_{\mathsf{c}}$ is adequate. Note that because $\sqsubseteq^{\mathsf{NA}}_{\mathsf{c}} \implies \sqsubseteq_{\mathsf{c}}$, we implicitly prove $\sqsubseteq_{\mathsf{c}}$ adequate. We define several versions of the abstractions with different levels

of context cutting:

$$\begin{aligned} B_1 \sqsubseteq^i_{\mathsf{c}} B_2 \;\triangleq\;\; & \forall X \in [\![B_1, \mathcal{A}]\!]^{\downarrow}_{vr}.\, \mathsf{cut}^i(X) \implies \\ & \exists Y \in [\![B_2, \mathcal{A}]\!]^{\downarrow}_{vr}.\, \mathsf{hist}_{\mathsf{E}}(X) \sqsubseteq_{\mathsf{E}} \mathsf{hist}_{\mathsf{E}}(Y) \wedge \\ & (\forall a.\, \mathsf{FinalNA}(X, a) \implies \exists b \in A(Y).\, a \leq^{X,Y}_{\mathsf{na}} b) \wedge \\ & (X \in [\![B_1, \mathcal{A}]\!]_{vr} \implies Y \in [\![B_2, \mathcal{A}]\!]_{vr}) \end{aligned}$$

We define several versions of the cutting predicate, incrementally cutting more of the context.

$$\mathsf{cut}^1(X) \triangleq \mathsf{cutR}(X)$$

$$\mathsf{cut}^2(X) \triangleq \mathsf{cutR}(X) \wedge \mathsf{cutW}(X)$$

$$\mathsf{cut}^3(X) \triangleq \mathsf{cutR}(X) \wedge \mathsf{cutW}(X) \wedge \mathsf{NAcutR}(X)$$

LEMMA 13 (ATOMIC READ CUTTING) $B_1 \sqsubseteq^1_{\mathsf{c}} B_2 \implies B_1 \sqsubseteq^{\mathsf{NA}}_{\mathsf{d}} B_2$

*Proof.*
- Pick an execution $X \in [\![B_1, \mathcal{A}, S]\!]^{\downarrow}_{d}$. We now want to build a corresponding execution such that $\mathsf{cutR}$ holds.
- Identify a subset $\mathcal{A}' \subseteq A(X)$ such that $\mathsf{cutR}(X|_{\mathcal{A}'})$ holds, and no larger subset exists. We call this maximal projected execution $X'$. We use $\mathcal{A}_R$ to refer to the removed actions $\mathcal{A} \setminus \mathcal{A}'$.
- It's straightforward to see that $\mathcal{A}_R \subseteq \mathsf{Read} \cap \mathsf{interf}(X)$. Context actions aren't required by the thread-local semantics, and removing context reads preserves validity, so $X' \in [\![B_1, \mathcal{A}, S]\!]^{\downarrow}_{d}$.
  It's also straightforward from the definition of $\mathsf{cutR}$ to see that any read $r$ in $\mathcal{A}_R$ is removed for one of two reasons:
  - *context-read.* The associated write for $r$ is in the context.
  - *duplicate-read.* The associated write is read by another context read $r'$ which is not removed. We call this $r'$ the *representative* for $r$.
- We have an execution $X' \in [\![L_1, \mathcal{A}, S]\!]^{\downarrow}_{d}$ such that $\mathsf{cutR}(X')$ holds. Now apply the assumption to produce an execution $\exists Y' \in [\![L_2, \mathcal{A}, S]\!]^{\downarrow}_{d}$ such that $\mathsf{hist}_{\mathsf{E}}(X') \sqsubseteq_{\mathsf{E}} \mathsf{hist}_{\mathsf{E}}(Y')$.
  Build a new execution $Y$ by re-injecting the actions from $\mathcal{A}_R$. As all of these actions are context reads, the only relation that must change is $\mathsf{rf}$.
  - If the action is a context-read, direct $\mathsf{rf}$ to the context write it pointed to in $X$. This must still exist by history inclusion.
  - If the action is a duplicate-read, direct $\mathsf{rf}$ to the write read by its representative. The origin for the representative write must exist by validity of $Y'$.

  It now remains to show that that $Y \in [\![L_2, \mathcal{A}, S]\!]^{\downarrow}_{d}$ and $\mathsf{hist}_{\mathsf{E}}(X) \sqsubseteq_{\mathsf{E}} \mathsf{hist}_{\mathsf{E}}(Y)$.
- To show that $Y \in [\![B_2, \mathcal{A}, S]\!]^{\downarrow}_{d}$, we only need to show that $Y$ is valid. Adding new atomic context reads to a valid execution is guaranteed to preserve validity, as long as they are equipped with valid origin writes in $\mathsf{rf}$.
- To show that $\mathsf{hist}_{\mathsf{E}}(X) \sqsubseteq_{\mathsf{E}} \mathsf{hist}_{\mathsf{E}}(Y)$, we have two obligations: $\mathsf{hbL}(Y) \subseteq \mathsf{hbL}(X)$, and $\mathsf{denyL}(Y) \subseteq \mathsf{denyL}(X)$. The former is a trivial consequence of the way we construct $Y$.
  For the latter, we reason by contradiction for each of the deny shapes:

- HBvsMO-d and Acyc-d: As context reads are terminal in hb, the only case we need to consider is the one where $u \in \mathcal{A}_R$ and the remainder of the shape is not removed. Otherwise the deny is entirely replicated in $Y'$, and thus in $X$. If $u$ is a duplicate-read, the deny is replicated in $Y$ using its representative. If $u$ is a context-read, a deny edge exists $w_1 \xrightarrow{\mathsf{d}} v$. In either case, it is easy to see that the deny $u \xrightarrow{\mathsf{d}} v$ must be replicable in $X$, contradicting the assumption.
- Cohere-d and Init-d: Similarly, the cases we need to consider are (1) $u \in \mathcal{A}_R$, (2) $v = r$ and $r \in \mathcal{A}_R$, and (3) both. In the first case, the same argument applies as with HBVSMO. In the second, we can replace $r$ with its representative. In both cases, it's straightforward to replicate the deny $u \xrightarrow{\mathsf{d}} v$ is replicated in $X$. The third case just combines the arguments from the other two.
- CoNA-d and CoNA-d2: Ruled out as actions in $\mathcal{A}_R$ must be hb-terminal. This precludes any such action participating in one of these non-atomic shapes.

– Finally, we need to show that any final NA action in $X$ is replicated in $Y$, and that $Y$ is complete if $X$ is complete. Both properties are inherited trivially from $Y'$.

LEMMA 14 (ATOMIC WRITE CUTTING) $B_1 \sqsubseteq^2_{\mathsf{c}} B_2 \implies B_1 \sqsubseteq^1_{\mathsf{c}} B_2$

*Proof.* – Pick an $X \in [\![B_1, \mathcal{A}, S]\!]^{\downarrow}_d$ such that $\mathsf{cutR}(X)$ holds.

– Now we build an $X'$ such that $X' \in [\![B_1, \mathcal{A}, S]\!]_d$ and $\mathsf{cutR}(X) \wedge \mathsf{cutW}'(X)$ holds. First identify the set of non-visible write actions for each location $z$:

$$\mathcal{A}^z = \{a \in \mathcal{A} \mid \mathsf{loc}(a) = z \wedge a \in (\mathsf{Write} \cap \mathsf{Atomic}) \wedge \neg\mathsf{visible}(a)\}$$

Partition this set into maximal disjoint nonempty subsets $\mathcal{B}^z_1$, $\mathcal{B}^z_2$ ... such that:

$$\mathcal{B}^z_i \subseteq \mathcal{A}^z \wedge (\forall a_1, a_2 \in \mathcal{B}^z_n.\, \neg\exists w \notin \mathcal{B}^z_n.\, a_1 \xrightarrow{\mathsf{mo}} w \xrightarrow{\mathsf{mo}} a_2)$$

In other words, each set $\mathcal{B}$ is a maximal set of non-visible writes so that there is no intervening write in mo. Thus, either a set $\mathcal{B}$ is mo-minimal / maximal, or it has a visible action which is its immediate mo-predecessor / successor. We call these actions $w^{\mathcal{B}}_p$ and $w^{\mathcal{B}}_s$ respectively. (The cases where $\mathcal{B}$ is minimal / maximal are ignored as they are simpler versions of the case where $w^{\mathcal{B}}_p$ and $w^{\mathcal{B}}_s$ exist.

Note that due to COHERENCE, if there is a LL-SC in $\mathcal{B}$, it must either read from a write in $\mathcal{B}$, or from $w^{\mathcal{B}}_p$. Similarly, if $w^{\mathcal{B}}_s$ is a LL-SC, it must read from a write in $\mathcal{B}$.

To build $X'$, replace each $\mathcal{B}$ with a single LL-SC pair $w^{\mathcal{B}}_n$ (as above, call this a *representative*). Take as the value that is read the value of $w^{\mathcal{B}}_p$, and take as the written value the mo-final value written in $\mathcal{B}$. We modify the rest of the execution as follows:

- As each set $\mathcal{B}$ is mo-contiguous in $X$, we don't need to modify mo other than to insert the new LL-SC pair.
- As the execution satisfies cutR, we have already kicked out all the context reads. We direct rf so that $w^{\mathcal{B}}_p \xrightarrow{\mathsf{rf}} w^{\mathcal{B}}_n$. If $w^{\mathcal{B}}_s$ is a LL-SC, we direct rf so that $w^{\mathcal{B}}_n \xrightarrow{\mathsf{rf}} w^{\mathcal{B}}_s$.
- Introducing $w^{\mathcal{B}}_n$ may generate new hb edges, so we regenerate hb according to HBDEF'.

– We now need to show that $X'$ is valid. This is simple for most of the axioms because the writes that are removed can only be read by their immediate mo-successor. However, if $w^{\mathcal{B}}_s$ is a LL-SC, then we might generate an hb-edge $w^{\mathcal{B}}_p \xrightarrow{\mathsf{hb}} w^{\mathcal{B}}_s$ which did not previously exist. We therefore need to show that ACYCLICITY, HBVSMO, COHERENCE, COHERNA still hold in $X'$.
  - HBVSMO, ACYCLICITY: The two writes $w_1, w_2$ responsible must be on a different location from $w^{\mathcal{B}}_p$ and $w^{\mathcal{B}}_s$: otherwise the violation would be an HBVSMO violation in $X$. Any hb-path between two actions on different locations must pass through the code. If the $w^{\mathcal{B}}_p$ and $w^{\mathcal{B}}_s$ are not themselves in the code, we can replicate the violation immediately using the hb-adjacent internal actions $a_p/a_s$.
  - COHERENCE, COHERNA: Again, the responsible writes / reads must be to a different location from $w^{\mathcal{B}}_p$ and $w^{\mathcal{B}}_s$. Otherwise we can generate a violation using the LL-SC $w^{\mathcal{B}}_s$, and the fact that mo follows hb. Apply the same reasoning as the previous point to replicate the violation in $X$.

  We also need to show that $\mathsf{cutR}(X') \wedge \mathsf{cutW}(X')$ holds. It's obvious that $\mathsf{cutR}(X')$ still holds – we have introduced no extra reads. $\mathsf{cutW}(X')$ holds because each new write $w^{\mathcal{B}}_n$ is separated in mo by a visible action.
– Apply the assumption to give an execution $Y' \in [\![L_2, \mathcal{A}, S]\!]$ such that $X' \sqsubseteq_{\mathsf{E}} Y'$. Now build the execution $Y$. To do this, replace each representative LL-SC $w^{\mathcal{B}}_n$ in $Y'$ with the corresponding actions in $\mathcal{B}$. In other words, for any pair of actions in a single set $\mathcal{B}$, take mo the same as in $\mathsf{mo}(X)$. For an action in $\mathcal{B}$ and some other action, relate it in mo as in $\mathsf{mo}(Y')$ for the set representative.
  We need to show that (1) $Y$ is valid, (2) $\mathsf{hist}_{\mathsf{E}}(X) \sqsubseteq_{\mathsf{E}} \mathsf{hist}_{\mathsf{E}}(Y)$.
– *Validity*. Modifying $Y'$ to $Y$ alters rf, mo, and hb.
  RFWF, RFVAL, MOWF, ATOM are obvious by construction.
  ACYCLICITY holds because hb edges are only removed between existing writes, and introduced between actions represented in $\mathcal{B}$, which are by definition unrelated to context actions aside from at $w^{\mathcal{B}}_p$ and $w^{\mathcal{B}}_s$. Therefore any cycle would exist inside $\mathcal{B}$, and thus in $X$.
  HBVSMO holds because actions in $\mathcal{B}$ are introduced at a single point in mo represented by $w^{\mathcal{B}}_n$. Any hb-edges inside $\mathcal{B}$ must be consistent with mo, or a similar violation could be replicated in $X$.
  COHERENCE, COHERNA holds because any violation for non-$\mathcal{B}$ reads/write could be replicated in $Y'$ using the representative LL-SC $w^{\mathcal{B}}_n$. A violation inside $\mathcal{B}$ could immediately be replicated in $X$.
  RFHBNA holds because any hb-path in $Y'$ that is broken in $Y$ must pass through the code. Therefore, the path must be replicable through sb, which contradicts the violation.
– $\mathsf{hbL}(Y) \subseteq \mathsf{hbL}(X)$. Actions in $\mathcal{B}$ in $X$ are only related to each other and $w^{\mathcal{B}}_p$ / $w^{\mathcal{B}}_s$ in hb. For paths in hb outside some $\mathcal{B}$, it must be that $\mathsf{hbL}(X) = \mathsf{hbL}(X') \subseteq \mathsf{hbL}(Y') = \mathsf{hbL}(Y)$. As paths inside $\mathcal{B}$ are identical between $X$ and $Y$, any hb-path can be replicated.
– $\mathsf{guarNA}(Y) \subseteq \mathsf{guarNA}(X)$. Trivial by the previous argument, and the fact $\mathcal{B}$-sets only cover atomic actions.

– $\mathsf{denyL}(Y) \subseteq \mathsf{denyL}(X)$. Prove by contradiction: assume a deny shape in $Y$ that is not in $X$.
  - HBvsMO-d / Acyc-d: Suppose a deny shape involving writes $w_1/w_2$.
    * $w_1/w_2$ not in any $\mathcal{B}$. As actions in $\mathcal{B}$ are not read/written in the code, any hb path which includes actions in $\mathcal{B}$ and which passes through the code, must enter and exit $\mathcal{B}$ through other context actions, $a, b$. There is a deny $a \xrightarrow{\mathsf{d}} b$ in $Y'$ by construction, and thus in $X$. hb-paths inside $\mathcal{B}$ are identical in $X$ and $Y$. Combining this gives us a deny in $X$.
    * $w_1/w_2$ entirely inside $\mathcal{B}$: reproducible trivially as mo/hb are identical between $Y$ and $X$.
    * $w_1$ outside $\mathcal{B}$, $w_2$ inside. In this case, there must be a deny in $Y'$ and $X'$ with the representative: $w_p^{\mathcal{B}} \xrightarrow{\mathsf{d}} b$ (using the same argument as above). Substituting $\mathcal{B}$ for the representative in $X$ builds the violation.
    * $w_2$ outside $\mathcal{B}$, $w_1$ inside. Symmetrical to previous case.
  - Cohere-d / Init-d: the deny shape involves writes $w_1/w_2/r$.
    * $w_1$, $w_2$, $r$ all in $\mathcal{B}$: replicated trivially.
    * $w_1$, $w_2$, $r$ all outside $\mathcal{B}$: replicated trivially.
    * $w_1$, $w_2$ in $\mathcal{B}$, $r$ outside: $r$ can only be $w_s^{\mathcal{B}}$, shape ruled out by construction.
    * $w_1$ in $\mathcal{B}$, $w_2$, $r$ outside: deny replicated in $Y'$ / $X'$ using representative. Rebuild the violation when reintroducing $\mathcal{B}$ in $X$.
    * All three outside $\mathcal{B}$: trivial.
    * $w_2$, $r$ in $\mathcal{B}$, $w_1$ outside: $w_1$ can only be $w_p^{\mathcal{B}}$, shape ruled out by construction.
    * $w_2$ in $\mathcal{B}$, $w_1$, $r$ outside: deny replicated in $Y'$ using representative, rebuild in $X$ when adding $\mathcal{B}$.
    * $r$ in $\mathcal{B}$, $w_1$, $w_2$ outside: $w_1$ can only be $w_p^{\mathcal{B}}$, shape ruled out by construction.
  - CoNA-d2: similar argument to Cohere-d, exept that some cases are ruled out by the fact that elements in $\mathcal{B}$ are necessarily atomic.

– $\mathsf{denyNA}(X) \subseteq \mathsf{denyNA}(Y)$. Similar argument to CoNA-d2 above.

– The Final NA and completeness properties are satisfied for the same reason as in the previous proof.

LEMMA 15 (NA READ CUTTING) $B_1 \sqsubseteq_{\mathsf{c}}^3 B_2 \implies B_1 \sqsubseteq_{\mathsf{c}}^2 B_2$

*Proof.* – Pick an $X \in [\![B_1, \mathcal{A}, S]\!]^{\downarrow}$ such that $\mathsf{cut}^2(X)$ holds. Build $X'$ using the same approach as in atomic read cutting: $X'$ is a maximal sub-execution such that $\mathsf{NAcutR}(X)$ holds.
From the structure of $\mathsf{NAcutR}$, the actions $\mathcal{A}_R$ removed from $X$ must all be non-atomic reads. Just as before, removed reads have a *representative* that remains in $X'$ and that reads from the same write. Unlike in the atomic case, reads from context writes also have representatives. This is necessary to detect new writes that might violated CoNA-d2 (which in turn is needed because NA writes are not ordered in mo).
$X'$ is valid because the axioms are invariant under read removal.

– We then apply the assumption to build an execution $Y' \in [\![B_2, \mathcal{A}, S]\!]_{vr}^{\downarrow}$. Finally we build $Y$ by restoring the cut actions, with $\mathsf{rf}(Y')$ built in the same way as for the atomic cutting case.

Almost-validity is preserved trivially because the inserted reads are not part of hb. Deny inclusion is ensured by the fact that the inserted reads are placed at the same position as their representatives: any violation would immediately be replicated by the representative. The FinalNA and completion property are unaffected from $Y'$.

LEMMA 16 (NA WRITE CUTTING) $B_1 \sqsubseteq_{\mathsf{c}} B_2 \implies B_1 \sqsubseteq^3_{\mathsf{c}} B_2$

*Proof.* – Pick an $X \in [\![B_1, \mathcal{A}, S]\!]^{\downarrow}$ such that $\mathsf{cut}^3(X)$ holds.

- As NAcutWdoesn't discriminate on the basis of mo, we can replace the set of all context writes to a location with a single representative write. We build $X'$ as a maximal sub-execution such that NAcutWholds, and 'orphan' context reads are removed. As NAcutRholds, each NA write has at most one context read. Note that as the execution is maximal, if at least one write to a location had an associated read, then the representative will have an associated read.
- Validity for $X'$ is trivial as the removed writes cannot participate in hb, or be read in the code. We then build $Y'$ by applying the theorem to give an almost-valid execution of $B_2$. Finally, we build $Y$ by re-inserting the removed reads and writes. The only relation that needs to be updated is rf, which associates removed reads with their origin write.
  Preservation of almost-validity follows from the fact that the inserted writes are disjoint from all other actions in the execution relations. Deny inclusion holds because any deny shape in $Y$ that involves a removed write / read can be easily replicated using the representative.

THEOREM 17 (CUT ADEQUACY) $B_1 \sqsubseteq_{\mathsf{c}} B_2 \implies B_1 \sqsubseteq^{\mathsf{NA}}_{\mathsf{d}} B_2$

*Proof.* Prove this as a sequence of implications:

$$B_1 \sqsubseteq_{\mathsf{c}} B_2 \implies B_1 \sqsubseteq^3_{\mathsf{c}} B_2 \implies B_1 \sqsubseteq^2_{\mathsf{c}} B_2 \implies B_1 \sqsubseteq^1_{\mathsf{c}} B_2 \implies B_1 \sqsubseteq^{\mathsf{NA}}_{\mathsf{d}} B_2$$

Each implication step is proved in a lemma above.

# F Proofs of Theorems 2 and 6 (full abstraction)

## F.1 Proof structure

We now sketch the proof structure for full abstraction, for simplicity eliding the treatment of non-atomics and LL-SC. The full proof is given in §F. Assume $B_1 \preccurlyeq_{\mathsf{bl}} B_2$; we have to prove $B_1 \sqsubseteq_{\mathsf{q}} B_2$.

1. Following the definition of $\sqsubseteq_{\mathsf{q}}$ (def. (7) in §4), consider arbitrary $\mathcal{A}$, $R$, and $X_1 \in [\![B_1, \mathcal{A}, R, \emptyset]\!]$ ($\emptyset$ is due to the fact that we ignore LL-SC).
2. We use $X_1$ to construct the special context $C_{X_1}$ (defined in §F.2). The context performs the actions specified by $\mathcal{A}$ and monitors executions to ensure that they do not significantly diverge from $X_1$, e.g., by checking that the values returned by context reads match those in $\mathcal{A}$. If $C_{X_1}$ detects a mismatch with $X_1$, it writes to a special observable error variable $e \in \mathsf{OVar}$. The context $C_{X_1}$ is constructed in such a way that for any code-block $B'$ and any execution $Y \in [\![C_{X_1}(B')]\!]$ in which $e$ is not written, the following three facts hold:

(a) the actions of $\mathcal{A}$ appear in $Y$, and the actions by $B'$ in $Y$ transform local variables in a way consistent with the call and ret actions in $X_1$;
(b) $\mathsf{hb}(Y)$ includes the edges in $R$;
(c) $\mathsf{hb}(Y)$ is included in the guarantee of $\mathsf{hist}(X_1)$.

3. We show that there is an execution $Z_1 \in [\![C_{X_1}(B_1)]\!]$ where the actions generated by $B_1$ match those in $X_1$, and where $e$ is not written; the latter implies that the above properties (a), (b) and (c) hold of $Z_1$.
4. Since $B_1 \preccurlyeq_{\mathsf{bl}} B_2$, by applying the definition of $\preccurlyeq_{\mathsf{bl}}$ (def. (2) in §4) to the special context $C_{X_1}$, we get an execution $Z_2 \in [\![C_{X_1}(B_2)]\!]$ where $e$ is never written.
5. By the construction of $C_{X_1}$, we know facts (a) and (b). Using this, we construct an execution $X_2 \in [\![B_2, \mathcal{A}, R, \emptyset]\!]$ where the actions generated by $B_2$ match those in $Z_2$ and the call and ret actions match those in $X_1$. Let $\mathsf{hist}(X_1) = (\mathcal{A}_1, G_1)$ and $X_2 = (\mathcal{A}_2, G_2)$. Using (a), we show $\mathcal{A}_1 = \mathcal{A}_2$ and using (c) we show $G_2 \subseteq G_1$. This establishes $\mathsf{hist}(X_1) \sqsubseteq_{\mathsf{h}} \mathsf{hist}(X_2)$, and by def. (7), gives $B_1 \sqsubseteq_{\mathsf{q}} B_2$.

## F.2 Context construction

We next describe the construction of the context $C_X$ for an execution $X \in [\![B, \mathcal{A}, R, \emptyset]\!]$ and argue that it satisfies the above properties (a)-(c). To illustrate the construction, we use the execution $X$ in Figure 4, for the block $B$ defined by def (5). The context $C_X$ is defined on the top of Figure 13 and an application to the example is given below (for brevity, we use syntactic sugar that elides manipulations of local variables).

The context $C_X$ is a parallel composition of threads: one for the parameter code-block $\{-\}$, and one each action in $\mathcal{A}$—these are collectively ranged over by $m$ in Figure 13. We introduce functions `call` and `ret` on the indices $m$, mapping $\{-\}$ to the call and ret actions in $X$, respectively, and acting as the identity otherwise. Recall that for our example execution $X$, the set $\mathcal{A}$ consists of the three writes outside the dashed rectangle. Our construction consists of several wrapper functions, introduced below.

1. Innermost is $\texttt{check}(m)$, which for brevity, we only describe informally. For a read or a write action $u \in \mathcal{A}$, $\texttt{check}(u)$ executes the corresponding operation and, in the case of a read, compares the value read with the one specified by $u$. The command $\texttt{check}(\{-\})$ initialises local variables to the values specified by the call action in $X$, runs the code-block, $\{-\}$, and then compares the local variables with the values specified by the ret action in $X$. If there is a mismatch in the above cases, `check` writes to the error variable $e$. In this way, it ensures that property (a) holds in error-free executions. We give an example of `check` on the right of Figure 13.
2. The wrappers $\texttt{Rrel}_m$ and $\texttt{Racq}_m$ ensure property (b). Recall the type (4) of $R$; in our running example from Figure 4, $R$ is given by the dashed edges. Each $\texttt{Rrel}_m$ is built up of a sequence of invocations of $\texttt{Rrel}_{u,v}$, one for each edge $(u, v) \in R$ outgoing from $u = \texttt{ret}(m)$; the wrapper $\texttt{Racq}_m$ is constructed symmetrically. These wrappers use watchdog variables $h_{u,v}$ to create happens-before edges as in the MP test of Figure 1. Namely, $\texttt{Rrel}_{u,v}$ and $\texttt{Racq}_{u,v}$ respectively write to and read from the variable $h_{u,v}$. If $\texttt{Racq}_{u,v}$ does *not* read the value written by $\texttt{Rrel}_{u,v}$, then it writes to the error variable $e$. The invocation of $\texttt{Rrel}_{u,v}$ is sequenced after $u$ and that of $\texttt{Racq}_{u,v}$ before $v$. Hence, any non-erroneous execution contains the shape

*Construction definition:*

$$C_X \triangleq \|_m \; (\mathtt{Racq}_m(\mathtt{Nrel}_m; \mathtt{check}(m)(\mathtt{Nacq}_m(\mathtt{Rrel}_m))))$$

$$\begin{aligned}
\mathtt{Rrel}_m &\triangleq \mathtt{Rrel}_{\mathtt{ret}(m),v_1}; \ldots; \mathtt{Rrel}_{\mathtt{ret}(m),v_n}, \\
&\quad \text{where } \{v_1,\ldots,v_n\} = \{v \mid (\mathtt{ret}(m),v) \in R\} \\
\mathtt{Rrel}_{u,v} &\triangleq \mathtt{store}(h_{u,v},1) \\
\mathtt{Racq}_m(N) &\triangleq \mathtt{Racq}_{u_1,\mathtt{call}(m)}(\ldots \mathtt{Racq}_{u_n,\mathtt{call}(m)}(N)\ldots), \\
&\quad \text{where } \{u_1,\ldots,u_n\} = \{u \mid (u,\mathtt{call}(m)) \in R\} \\
\mathtt{Racq}_{u,v}(N) &\triangleq \mathtt{if}\ (\mathtt{load}(h_{u,v}))\ N\ \mathtt{else}\ \mathtt{store}(e,1) \\
\mathtt{Nrel}_m &\triangleq \mathtt{Nrel}_{\mathtt{call}(m),v_1}; \ldots; \mathtt{Nrel}_{\mathtt{call}(m),v_n}, \\
&\quad \text{where } \{v_1,\ldots,v_n\} = \{v \mid (\mathtt{call}(m),v) \in H\} \\
\mathtt{Nrel}_{u,v} &\triangleq \mathtt{store}(g_{u,v}) \\
\mathtt{Nacq}_m(N) &\triangleq \mathtt{Nacq}_{u_1,\mathtt{ret}(m)}(\ldots \mathtt{Nacq}_{u_n,\mathtt{ret}(m)}(N)\ldots), \\
&\quad \text{where } \{u_1,\ldots,u_n\} = \{u \mid (u,\mathtt{ret}(m)) \in H\} \\
\mathtt{Nacq}_{u,v}(N) &\triangleq \mathtt{if}\ (\neg\mathtt{load}(g_{u,v}))\ N\ \mathtt{else}\ \mathtt{store}(e,1) \\
\mathtt{check}(\{-\}) &\triangleq \mathtt{l1} := 0; \mathtt{l2} := 0; \\
&\quad \{-\}; \\
&\quad \mathtt{if}\ (\mathtt{l1} \neq 1)\ \{\mathtt{store(e,1)}\}; \\
&\quad \mathtt{if}\ (\mathtt{l2} \neq 1)\ \{\mathtt{store(e,1)}\}
\end{aligned}$$

---

*Example application:*

$$\begin{aligned}
C_X = \quad & \mathtt{Racq}_{\mathsf{store(x,2),call}}(\mathtt{Nrel}_{\mathsf{call,store(x,1)}}; \mathtt{Nrel}_{\mathsf{call,store(f,1)}}; \mathtt{check}(\{-\})) \\
\| \; & \mathtt{store(x,2)}; \mathtt{Rrel}_{\mathsf{store(x,2),ret}}; \mathtt{Rrel}_{\mathsf{store(x,2),store(x,2)}} \\
\| \; & \mathtt{Racq}_{\mathsf{store(x,2),store(x,1)}}(\mathtt{store(x,1)}; \mathtt{Nacq}_{\mathsf{ret,store(x,1)}}(\mathtt{Rrel}_{\mathsf{store(x,1),store(f,1)}})) \\
\| \; & \mathtt{Racq}_{\mathsf{store(x,1),store(f,1)}}(\mathtt{store(f,1)}; \mathtt{Nacq}_{\mathsf{ret,store(f,1)}}(\mathtt{skip}))
\end{aligned}$$

**Fig. 13.** *Top:* Definition of the construction of $C_X$ for $X \in [\![B,\mathcal{A},R]\!]$. We define $H$ and check() in the text. The symbol $m$ ranges over context actions $\mathcal{A}$ and a hole $\{-\}$. *Bottom:* Example of check() and the construction for the execution in Figure 4.

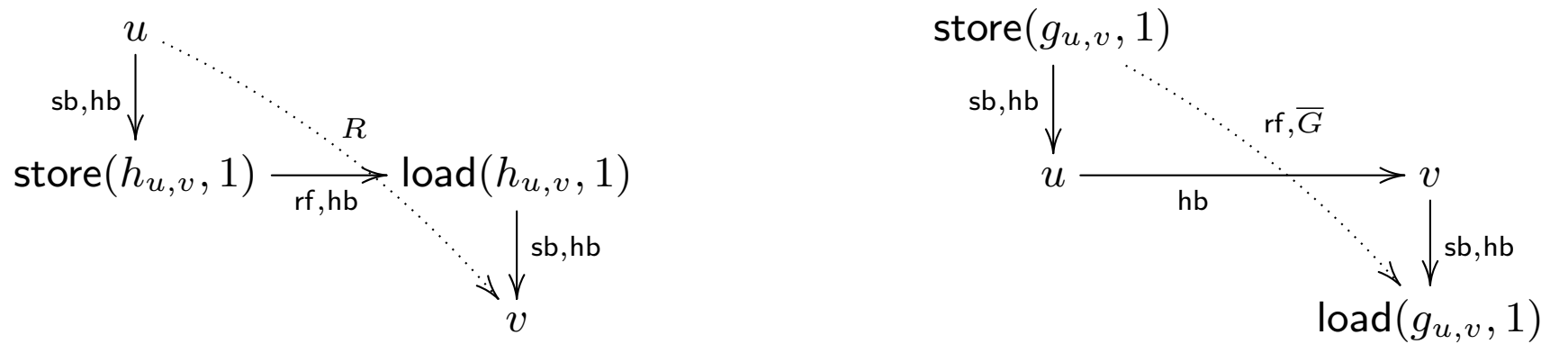

**Fig. 14.** Context construction execution shapes. *Left:* Shape enforcing edges in $R$. *Right:* shape prohibiting edges not in $G$.

on the left of Figure 14. This reproduces the required $R$ edge $(u, v)$ in the happens-before. In our running example, the edge $(\mathsf{store}(\mathtt{x}, 2), \mathsf{call}) \in R$ is reproduced by the invocation of $\mathtt{Racq}_{\mathsf{store}(\mathtt{x},2),\mathsf{call}}$ on the first thread and $\mathtt{Rrel}_{\mathsf{store}(\mathtt{x},2),\mathsf{call}}$ on the second.

3. The wrappers $\mathtt{Nrel}_m$ and $\mathtt{Nacq}_m$ ensure property (c), prohibiting new happens-before edges beyond those in the original guarantee $G$ of $\mathsf{hist}(X)$. We identify pairs that must be monitored with the relation $H$: the edges of $\overline{G}$ matching the type (6) of a guarantee that are not already covered by the reverse of $R$. In our running example from Figure 4, the edges from $\overline{G}$ that we need to consider are $(\mathsf{call}, \mathsf{write}(\mathtt{f}, 1))$ and $(\mathsf{call}, \mathsf{write}(x, 1))$. Each $\mathtt{Nrel}_m$ is built up of a sequence of invocations of $\mathtt{Nrel}_{u,v}$, one for each edge $(u, v) \in H$ outgoing from $u = \mathtt{call}(m)$; the wrapper $\mathtt{Racq}_m$ is constructed symmetrically. The above wrappers detect errant happens-before edges using watchdog variables $g_{u,v}$, again relying on the mechanics of the MP test of Figure 1. Namely, $\mathtt{Nrel}_{u,v}$ and $\mathtt{Nacq}_{u,v}$ respectively write to and read from a watchdog variable $g_{u,v}$. If $\mathtt{Nacq}_{v,u}$ *does* read the value written by $\mathtt{Nrel}_{v,u}$, then it writes to the error variable $e$. The invocation of $\mathtt{Rrel}_{u,v}$ is sequenced *before* $u$ and that of $\mathtt{Racq}_{u,v}$ *after* $v$. Hence, if an execution includes a happens-before edge $(u, v)$, then it contains the shape shown on the right of Figure 14 (omitting the write to the error location). Here the happens-before edge $(u, v)$ and the RFVAL axiom (§3) force the read in $\mathtt{Nacq}_{u,v}$ to read from the write in $\mathtt{Nrel}_{u,v}$, leading to a write to $e$. Hence, a non-erroneous execution does not contain errant happens-before edges. In our example the edge $(\mathsf{call}, \mathsf{store}(\mathtt{f}, 1)) \in \overline{G}$ is covered by the invocation of $\mathtt{Nrel}_{\mathsf{call},\mathsf{store}(\mathtt{f},1)}$ on the first thread and $\mathtt{Nacq}_{\mathsf{call},\mathsf{store}(\mathtt{f},1)}$ on the fourth.

*Context construction* The full context construction including LL/SC is included in Figure 15. The key difference from Figure 13 is that successful context LL/SC pairs in $X$ are arranged on a single thread, allowing the store conditional to suceed in $C_X$.

*Proof of Theorem 6.* We now prove Theorem 6: full abstraction of $\sqsubseteq_{\mathsf{q}}^{\mathsf{NA}}$ for programs and contexts that do not use read-modify-write accesses, $B_1 \preccurlyeq_{\mathsf{bl}}^{\mathsf{NA}} B_2 \implies B_1 \sqsubseteq_{\mathsf{q}}^{\mathsf{NA}} B_2$. Note that this implies Theorem 2.

*Proof.* – Start by choosing an arbitrary $R$, $S$ and $X \in [\![B_1, \mathcal{A}, R, S]\!]_v^{\mathsf{NA}}$. It remains to show that:

$$\begin{aligned}&\exists Y \in [\![B_2, \mathcal{A}, R, S]\!]_v^{\mathsf{NA}}.\\&(\mathsf{safe}(Y) \implies \mathsf{safe}(X) \wedge \mathsf{hist}(X) \sqsubseteq_{\mathsf{h}} \mathsf{hist}(Y)) \wedge\\&(\neg\mathsf{safe}(Y) \implies \exists X_1' \in (X)^{\downarrow}. \exists X_2' \in (Y)^{\downarrow}. \neg\mathsf{safe}(X_2') \wedge \mathsf{hist}(X_1') \sqsubseteq_{\mathsf{h}} \mathsf{hist}(X_2'))\end{aligned} \tag{11}$$

$$
\begin{aligned}
C_X &= (;_{(m_1,m_2)\in\mathsf{at}}\, a(m_1); a(m_2)) \parallel (\|_{m\setminus|\mathsf{at}|}\, a(m)) \\
a(m) &= \mathtt{Racq}_m(\mathtt{Nrel}_m; \mathtt{check}(m)(\mathtt{Nacq}_m(\mathtt{Rrel}_m))) \\
\mathtt{Rrel}_m &= \mathtt{Rrel}_{\mathtt{ret}(m),v_1}; \ldots; \mathtt{Rrel}_{\mathtt{ret}(m),v_n}, \\
&\quad \text{where } \{v_1,\ldots,v_n\} = \{v \mid (\mathtt{ret}(m), v) \in R\} \\
\mathtt{Rrel}_{u,v} &= \mathtt{store}(h_{u,v}, 1) \\
\mathtt{Racq}_m(N) &= \mathtt{Racq}_{u_1,\mathtt{call}(m)}(\ldots \mathtt{Racq}_{u_n,\mathtt{call}(m)}(N)\ldots), \\
&\quad \text{where } \{u_1,\ldots,u_n\} = \{u \mid (u, \mathtt{call}(m)) \in R\} \\
\mathtt{Racq}_{u,v}(N) &= \mathtt{if}\ (\mathtt{load}(h_{u,v}))\ N\ \mathtt{else}\ \mathtt{store}(e,1) \\
\mathtt{Nrel}_m &= \mathtt{Nrel}_{\mathtt{call}(m),v_1}; \ldots; \mathtt{Nrel}_{\mathtt{call}(m),v_n}, \\
&\quad \text{where } \{v_1,\ldots,v_n\} = \{v \mid (\mathtt{call}(m), v) \in H\} \\
\mathtt{Nrel}_{u,v} &= \mathtt{store}(g_{u,v}) \\
\mathtt{Nacq}_m(N) &= \mathtt{Nacq}_{u_1,\mathtt{ret}(m)}(\ldots \mathtt{Nacq}_{u_n,\mathtt{ret}(m)}(N)\ldots), \\
&\quad \text{where } \{u_1,\ldots,u_n\} = \{u \mid (u, \mathtt{ret}(m)) \in H\} \\
\mathtt{Nacq}_{u,v}(N) &= \mathtt{if}\ (\neg\mathtt{load}(g_{u,v}))\ N\ \mathtt{else}\ \mathtt{store}(e,1)
\end{aligned}
$$

**Fig. 15.** Context construction: $m$ ranges over context actions $\mathcal{A}$ and a code-block $B'$.

- Apply the construction lemma (Lemma 18, below) to $B_1$, $R$, $S$ and $X$ to find a context $C_X$, and an execution $Z$:

$$
\begin{aligned}
&Z \in [\![C_X(B_1)]\!]^{\mathsf{NA}}_v \land \\
&\mathsf{code}(Z) = X \land \\
&\mathsf{hbC}(Z) = R \land \mathsf{atC}(Z) = S \land \\
&\forall B'. \forall Z' \in [\![C_X(B')]\!]^{\mathsf{NA}}_v. \\
&\quad ((A(\mathsf{contx}(Z)) = A(\mathsf{contx}(Z'))) \implies \\
&\qquad (\mathsf{hist}(Z) \sqsubseteq_{\mathsf{h}} \mathsf{hist}(Z')) \land \mathsf{at}(\mathsf{contx}(Z)) = \mathsf{at}(\mathsf{contx}(Z')))\land \\
&\quad (\neg\mathsf{safe}(Z') \implies \\
&\qquad \exists X' \in X^{\downarrow}. \exists W \in ([\![B', \mathcal{A}, R, S]\!]^{\mathsf{NA}}_v)^{\downarrow}. \neg\mathsf{safe}(W) \land \mathsf{hist}(X') \sqsubseteq_{\mathsf{h}} \mathsf{hist}(W))
\end{aligned} \tag{12}
$$

- Specialise observation with the context $C_X$ and the set of all variables used in $C_X$, $V_{C_X}$, to get:

$$
\begin{aligned}
(\quad &\neg\mathsf{safe}(C_X(B_2)) \lor \\
&(\mathsf{safe}(C_X(B_1)) \land \\
&\quad \forall X \in [\![C_X(B_1)]\!]^{\mathsf{NA}}_v. \exists Y \in [\![C_X(B_2)]\!]^{\mathsf{NA}}_v. \\
&\qquad \mathcal{A}(X|_{V_{C_X}}) = \mathcal{A}(Y|_{V_{C_X}}))) \land \\
&\qquad \mathsf{hb}(X|_{V_{C_X}}) \subseteq \mathsf{hb}(Y|_{V_{C_X}})))
\end{aligned} \tag{13}
$$

and then case split on $\mathsf{safe}(C_X(B_2))$.

- **Case 1:** $\mathsf{safe}(C_X(B_2))$.
  - By 13, there is an execution, $Z'$ of $C_X(B_2)$ with:

$$
\mathsf{hb}(Z|_{V_{C_X}}) = \mathsf{hb}(Z'|_{V_{C_X}}) \land \mathcal{A}(Z|_{V_{C_X}}) = \mathcal{A}(Z'|_{V_{C_X}})
$$

- By construction of $C_X$, the variables $V_{C_X}$ cover all context variables, so we have:

$$\mathsf{hb}(\mathsf{contx}(Z)) = \mathsf{hb}(\mathsf{contx}(Z')) \wedge A(\mathsf{contx}(Z)) = A(\mathsf{contx}(Z'))$$

- Appealing to 12, we have:

$$\mathsf{hist}(Z) \sqsubseteq_{\mathsf{h}} \mathsf{hist}(Z')$$

- Now apply the decomposition lemma to $Z'$ to get the execution $Y$:

$$Y \in [\![B_2, \mathcal{A}, \mathsf{hbC}(Z'), \mathsf{atC}(Z')]\!]_v^{\mathsf{NA}} \wedge \mathsf{code}(Z') = Y$$

- Now simplify using the definition of $\mathsf{hbC}$ and $\mathsf{atC}$.

$$Y \in [\![B_2, \mathcal{A}, R, S]\!]_v^{\mathsf{NA}} \wedge \mathsf{hbC}(Z') = R = \mathsf{hbC}(Z) \wedge \mathsf{atC}(Z') = S = \mathsf{atC}(Z)$$

- Choose $Y$ as the witness for our goal 11. Note that the presence of a safety violation in $X$ or $Y$ would contradict the safety of $C_X(B_2)$ and $C_X(B_1)$. It is left to show that:

$$\mathsf{hist}(X) \sqsubseteq_{\mathsf{h}} \mathsf{hist}(Y)$$

- Unfolding the definition of $\sqsubseteq_{\mathsf{h}}$, we have:

$$A(Z) = A(Z') \wedge \mathsf{hbL}(Z') \subseteq \mathsf{hbL}(Z)$$

and it is left to show that,

$$A(X) = A(Y) \wedge \mathsf{hbL}(Y) \subseteq \mathsf{hbL}(X)$$

- Note that $X$ and $Y$ are the code-block projections of $Z$ and $Z'$ respectively, and we are done.

– **Case 2:** $\neg\mathsf{safe}(C_X(B_2))$.

- Identify an unsafe valid execution of $C_X(B_2)$, $Z'$, and specise the final conjunct of 12 with $B_2$ and $Z'$ to get:

$$X' \in X^{\downarrow} \wedge W \in ([\![B_2, \mathcal{A}, R, S]\!]_v^{\mathsf{NA}})^{\downarrow} \wedge \neg\mathsf{safe}(W) \wedge \mathsf{hist}(X')^{\downarrow} \sqsubseteq_{\mathsf{h}} \mathsf{hist}(W)$$

Then by the definition of ${}^{\downarrow}$, there exists a $W' \in [\![B_2, \mathcal{A}, R, S]\!]_v^{\mathsf{NA}}$ such that $W \in W'^{\downarrow}$, and this case is completed by noting that $W'$ and $W$ satisfy 11.

### F.3 Context construction

LEMMA 18 (CONSTRUCTION LEMMA)

$$
\begin{aligned}
&\forall B\, \mathcal{A}, R, S\, .\forall X \in [\![B, \mathcal{A}, R, S]\!]_v^{\mathsf{NA}}. \\
&\quad \exists C_X. \exists Z \in [\![C_X(B)]\!]_v^{\mathsf{NA}}. \\
&\qquad (\mathsf{code}(Z) = X) \wedge \\
&\qquad (\mathsf{hbC}(Z) = R) \wedge (\mathsf{atC}(Z) = S) \wedge \\
&\qquad \forall B'. \forall Z' \in [\![C_X(B')]\!]_v^{\mathsf{NA}}. \\
&\qquad\quad ((A(\mathsf{contx}(Z)) = A(\mathsf{contx}(Z'))) \implies \\
&\qquad\qquad \mathsf{hist}(Z) \sqsubseteq_{\mathsf{h}} \mathsf{hist}(Z') \wedge \mathsf{at}(\mathsf{contx}(Z)) = \mathsf{at}(\mathsf{contx}(Z'))) \wedge \\
&\qquad\quad (\neg \mathsf{safe}(Z') \implies \\
&\qquad\qquad \exists X' \in X^{\downarrow}. \exists W \in ([\![B', \mathcal{A}, R, S]\!]_v^{\mathsf{NA}})^{\downarrow}. \neg \mathsf{safe}(W) \wedge \mathsf{hist}(X') \sqsubseteq_{\mathsf{h}} \mathsf{hist}(W))
\end{aligned}
$$

*Proof.* – Start by choosing an arbitrary $B$, $\mathcal{A}$, $R$, $S$ and $X \in [\![B, \mathcal{A}, R, S]\!]$.

- Construct the client $C_X$ as specified in Figure 13 with one minor change: have `check` halt the thread if the error variable is written.
  It remains to show that there exists $Z \in [\![C_X(B)]\!]_v^{\mathsf{NA}}$ such that:
  1. $(\mathsf{code}(Z) = X) \wedge (\mathsf{hbC}(Z) = R) \wedge (\mathsf{atC}(Z) = S)$
  2. $\forall B'. \forall Z' \in [\![C_X(B')]\!]_v^{\mathsf{NA}}.$
     $((A(\mathsf{contx}(Z)) = A(\mathsf{contx}(Z'))) \implies \mathsf{hist}(Z) \sqsubseteq_{\mathsf{h}} \mathsf{hist}(Z') \wedge \mathsf{at}(\mathsf{contx}(Z)) = \mathsf{at}(\mathsf{contx}(Z'))) \wedge$
     $(\neg \mathsf{safe}(Z') \implies \exists X' \in X^{\downarrow}. \exists W \in ([\![B', \mathcal{A}, R, S]\!]_v^{\mathsf{NA}})^{\downarrow}. \neg \mathsf{safe}(W) \wedge \mathsf{hist}(X') \sqsubseteq_{\mathsf{h}} \mathsf{hist}(W))$
- We first establish 1.
  - Appealing to the thread local semantics and the structure of $C_X(B)$, choose $Z_p$, a pre-execution of $C_X(B)$ that does not write the error variable, and whose code projection matches $X$.
  - Note that at is generated from the thread-local semantics matching $X$, and for the context part, each LL/SC pair is in its own thread, so there is only one way to link them.
  - Construct $mo$ as follows: for code actions choose these edges to match $X$, and for the context part, note that there is no choice: at each location there is only one write after the initialisation.
  - Construct rf as follows: for code actions choose these edges to match $X$, and for that context actions set rf to be coincident with an $R$ edge in the case of Racq or from the initialisation write in the case of Nacq. Note that the context projection of happens-before matches $R$ by construction.
  - Let $Z$ be the combination of $Z_p$, mo and rf. Show that $Z$ is valid.
    * The only axioms that could fail are *Acyclicity* over some Racq or *Coherence* over some Nacq.
    * In the first case, any cycle would be made up of code hb and $R$ edges, and would also be a cycle in $X$, a contradiction.
    * A *Coherence* violation over some Nacq implies the existence of a hb edge from the associated Nrel to the Nacq. This violates the rules used to construct $C_X$, and is a contradiction.
- Now establish 2.

- Start by choosing arbitrary $B'$ and $Z' \in [\![C_X(B')]\!]_v^{\mathsf{NA}}$.
- First, show that $(A(\mathsf{contx}(Z)) = A(\mathsf{contx}(Z'))) \implies \mathsf{hist}(Z) \sqsubseteq_{\mathsf{h}} \mathsf{hist}(Z') \wedge \mathsf{at}(\mathsf{contx}(Z)) = \mathsf{at}(\mathsf{contx}(Z'))$
  - ∗ $Z$ does not write $e$, so neither does $Z'$ (they have an equal context projection).
  - ∗ By construction of $C_X$, the histories of $Z$ abstract the histories of $Z'$ and the at relations match.
- Now suppose $Z'$ is unsafe. It remains to show:

$$\exists X' \in X^{\downarrow}.\, \exists W \in ([\![B', \mathcal{A}, R, S]\!]_v^{\mathsf{NA}})^{\downarrow}.\, \neg\mathsf{safe}(W) \wedge \mathsf{hist}(X') \sqsubseteq_{\mathsf{h}} \mathsf{hist}(W)$$

  - ∗ $C_X$ uses only atomic and local variables that cannot exhibit safety violations: each violation must be amongst the actions of $\mathcal{A}$ and the actions generated by $B'$.
  - ∗ Identify a safety violation in $Z'$ and consider the prefix $Z'_p$ containing only $\mathsf{hb} \cup \mathsf{rf}$ predecessors of the actions of the violation.
  - ∗ There are no writes to $e$ in $Z'_p$: after any such write, the thread is stopped, so it cannot appear in the prefix $Z'_p$.
  - ∗ Below, we establish that for every thread of $Z'_p$ except those that contain the safety violation from which it is constructed, the error variable is not written in the corresponding thread of $Z'$.
    - · Consider the $\mathsf{hb} \cup \mathsf{rf}$ edges that draw actions into the prefix $Z'_p$ from some thread $t$, there are two cases: the edge arises from a `Rrel`/`Racq` pair, or it is created by a read, from write $w$, in the code block or a context action.
    - · In the first case $e$ is not written in `check` or `Nacq` on $t$, because that would halt the thread before the call to `Rrel`.
    - · In the second case, no call to `Nacq` writes $e$, and calls of `commit` only write to $e$ in the case of a failing store conditional, contradicting the existence of write $w$ in $Z'$, so $w$ is only performed in threads that never write to $e$.
  - ∗ From $Z'_p$, we construct $W$ by applying the decomposition lemma $Z'_p$ to get an execution $W$, completing this to an execution in $[\![B', \mathcal{A}, R, S]\!]_v^{\mathsf{NA}}$, and observing that $W$ is in $([\![B', \mathcal{A}, R, S]\!]_v^{\mathsf{NA}})^{\downarrow}$. $W$ is unsafe by construction.
  - ∗ Take $A$ to be the set of all context actions in $W$ together with the call and ret actions present in $W$. Let $X'$ be the projection of $X$ to the $\mathsf{hb} \cup \mathsf{rf}$ predecessors of $A$, so that $X' \in X^{\downarrow}$.
  - ∗ It remains to show that there is no edge in $\mathsf{hb}(W)$ between the actions of $A$ that is not present in the guarantee of $X'_p$, $G'$. By construction of $W$, any extraneous $\mathsf{hb}(W)$ edge must end at one of the threads hosting the violation. There is only one code block, so at least one of the threads has to be executing a context action. There is no single action that both creates an incoming hb edge and causes a safety violation, so any additional $\mathsf{hb}(W)$ edge must end at the code block. In that case, $W$ does not include the ret action following the racy action in hb, and neither does $X'$, and there can be no new edge in $G'$.